\documentclass{optica-article}

\journal{opticajournal} 

\articletype{Research Article}

\usepackage{lineno}
\linenumbers 

\usepackage[english]{babel}
\usepackage[utf8]{inputenc}
\usepackage[T1]{fontenc}
\usepackage{graphicx}
\usepackage{subcaption}
\graphicspath{ {./images/} }
\usepackage{mathtools}
\usepackage{physics}
\usepackage{upgreek}
\usepackage{footmisc}
\usepackage{booktabs}
\usepackage{array}
\usepackage{siunitx}
\usepackage[table]{xcolor}
\definecolor{lightgrey}{RGB}{230,230,230}

\begin{document}
\nolinenumbers
\title{Systematic Study of Amorphous ABC Heterostructures at the Atomic Scale as a Second-Order Nonlinear Optical Metamaterial}

\author{Martin Mičulka,\authormark{1,2} Jinsong Liu,\authormark{1} Sebastian Beer,\authormark{1} Raihan Rafi,\authormark{1} Denys Sevriukov,\authormark{1,3} Sergiy Yulin,\authormark{3} Vladimir Roddatis,\authormark{4} Stephan Gierth,\authormark{5} Stefan Nolte,\authormark{1,3}  Sven Schröder,\authormark{3} Isabelle Staude,\authormark{6} Andreas Tünnermann,\authormark{1,3} and Adriana Szeghalmi\authormark{1,3,*}}

\address{\authormark{1}Friedrich Schiller University Jena, Faculty of Physics and Astronomy, Institute of Applied Physics, Albert-Einstein-Str. 15, 07745 Jena, Germany\\
\authormark{2}Australian National University, Research School of Physics, Canberra, ACT, 2601, Australia\\
\authormark{3}Fraunhofer Institute for Applied Optics and Precision Engineering IOF, Albert-Einstein-Str. 7, 07745 Jena, Germany\\
\authormark{4}GFZ Helmholtz Centre for Geosciences, Telegrafenberg, 14473 Potsdam, Germany, ORCID: 0000-0002-9584-0\\
\authormark{5}Fraunhofer Institute for Microstructure of Materials and Systems IMWS,
Walter-Hülse-Straße 1,
06120 Halle (Saale), Germany\\
\authormark{6}Friedrich Schiller University Jena, Faculty of Physics and Astronomy, Institute of Solid State Physics, Max-Wien-Platz 1, 07743 Jena, Germany}

\email{\authormark{*}Adriana.Szeghalmi@iof.fraunhofer.de} 


\begin{abstract*} 
Systematic exploration of amorphous ABC heterostructures revealed that nanoscale morphological modifications markedly improved their artificial bulk second-order susceptibility. These amorphous birefringent heterostructures were fabricated through plasma-enhanced atomic layer deposition of three oxides, effectively breaking the centrosymmetry on the nanoscale. We observe a dependence of the optical nonlinearity on the thickness variation of three constituent materials, SiO$_2$, TiO$_2$, and Al$_2$O$_3$, ranging from tens of nanometers to the atomic scale, and these thin films exhibit second-order susceptibility at their interfaces. Our findings reveal that the enhancement of nonlinear optical properties is strongly correlated with a high density of layers and superior interface quality, where the interface second-order nonlinearity transitions to bulk-like second-harmonic generation. An effective bulk second-order susceptibility of $\chi_{zzz} = 2.0 \pm 0.2$~pm/V at the wavelength of 1032~nm is achieved, comparable to typical values for conventional monocrystalline nonlinear materials.

\end{abstract*}

\section{Introduction}
\begin{figure}[htbp]
\centering\includegraphics[width=7cm]{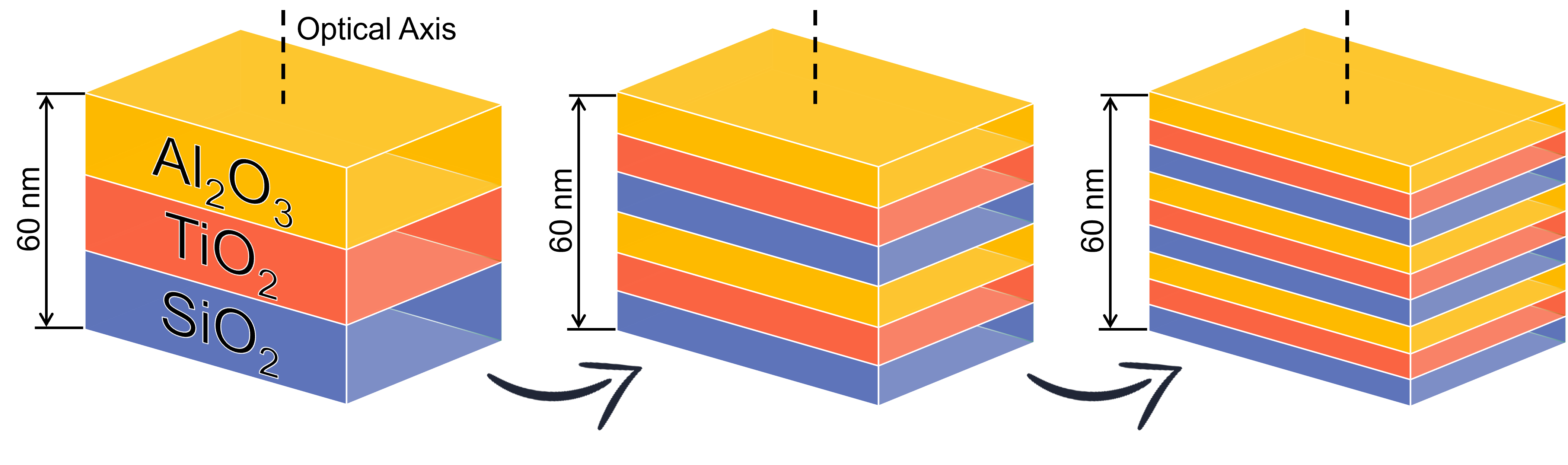}
\caption{The visualization of the ABC-type heterostructures with an optical axis perpendicular to the layers. Changing the density of layers influences their nonlinear optical properties.}
\label{fig:ABC}
\end{figure}

The rise of atomic layer deposition (ALD) has revolutionized numerous technologies, including transistors~\cite{ALD-tranzistor}, memory devices~\cite{ALD-memory}, solar cells~\cite{solarcell}, catalysts~\cite{catalyst}, batteries~\cite{batteries}, and conformal optical coatings~\cite{AdrianaALD}. However, its impact on the field of nonlinear optics remains largely untapped~\cite{ABC-Alloatti, ABC-Clemmen}. Here, we investigate the potential of ALD-deposited amorphous heterostructures, specifically ABC-type nanolaminates consisting of three distinct dielectrics SiO$_2$ (A), TiO$_2$ (B), and Al$_2$O$_3$ (C) as shown in Fig.~\ref{fig:ABC}, to create CMOS-compatible second-order nonlinear metamaterials. The core concept of metamaterials is to engineer materials with properties that are not found in nature by designing their structure at a scale smaller than the wavelength of light they interact with. Atomic layer deposition enables unparalleled precision in the fabrication of the thin films, building one atomic layer at a time with exceptional control and uniformity. Previous studies on layered composites with nonzero bulk second-order susceptibility assumed that only non-centrosymmetric constituents contribute to nonlinearity. However, they neglected interface contributions to the nonlinear susceptibility of the heterostructure~\cite{Boyd:94}. Our goal is to leverage the often overlooked interface, or surface, second-order nonlinearity~\cite{heinz1991second, shen1989optical} by stacking three thin oxide layers in 2D heterostructures, thus significantly enhancing the effective nonlinear response by breaking the centrosymmetry as discussed by L.~Alloatti~et~al. and S.~Clemmen~at~al.~\cite{ABC-Alloatti, ABC-Clemmen}. Impressively, these heterostructures exhibit a robust, bulk-like nonlinear response, even though all the constituent oxides lack inherent bulk second-order nonlinearity and are amorphous. This produces a highly distinctive material where second-order nonlinear processes arise not from bulk nonlinearity but from surface nonlinearity at the interfaces between centrosymmetric constituents. 

Previous studies on ternary oxide heterostructures (\text{Al$_2$O$_3$}/\text{TiO$_2$} and \text{SiO$_2$}/\text{HfO$_2$}) have demonstrated the superior structural and optical properties of ALD-grown thin films~\cite{palabi, PallabiHfO2}. These structures offer temperature and chemical stability, low-temperature growth requirements of $100~^{\circ}\text{C}$, and conformal deposition on a wide range of nano- and micro-structured substrates, without the need to account for substrate compatibility or crystal growth orientation, unlike nonlinear III–V semiconductors such as GaAs~\cite{GaAs}. Moreover, they bypass the high-temperature manufacturing requirements of lithium niobate on insulator (LNOI), another widely used nonlinear platform~\cite{LNOI}. They also provide the flexibility to engineer the effective refractive index for specific applications by adjusting the ratio of the constituent oxides~\cite{palabi}. Oxide heterostructures exhibit outstanding optical transparency across the visible spectrum to UV spectral range, surpassing the performance of typical III-V technologies. Moreover, the ALD deposition technique and dielectrics used are standard in the semiconductor industry.
Altogether, these advantages make ABC heterostructures a powerful platform for integrating photonic structures on a chip using only the commonly available oxides and established semiconductor processes.

In bulk media with inversion symmetry, second-order nonlinear processes are forbidden under the electric dipole approximation, leading to zero bulk second-order susceptibility, $\chi_{\text{bulk}}^{(2)} = 0$~\cite{Shen, Boyd1992}, and no second harmonic enhancement is observed with varying thickness in pure samples of centrosymmetric materials A, B, or C. However, at the interface between two centrosymmetric media, the inversion symmetry, defined by the transformation $(x, y, z) \to (-x, -y, -z)$, is broken. This symmetry breaking results in a nonzero surface second-order susceptibility, $\chi_{\text{s}}^{(2)} \neq 0$, enabling electric-dipole-allowed second-harmonic generation (SHG) at surfaces and interfaces~\cite{Shen1989, GuyotSionnest1987}. ABC heterostructures make use of this principle by stacking nanometer-thick amorphous layers of three centrosymmetric materials A, B, and C. The inversion symmetry is broken locally at each AB, BC, and CA interface. In the case of AB stacks, the centrosymmetry is broken only locally  and not globally. However, in ABC stacks, the inversion symmetry is broken both locally at the interfaces and globally, prohibiting total destructive interference of the second harmonic, resulting in a net $\chi_{\text{s}}^{\text{ABC}}$ for the ABC stack. The literature suggests that ABCD stacks, such as those incorporating SiO$_2$/TiO$_2$/Al$_2$O$_3$ with the addition of HfO$_2$, may outperform ABC stacks. However, this potential enhancement remains experimentally unverified~\cite{ABC-Alloatti, ABC-Clemmen}. Unfortunately, clear studies are lacking on which combination of inversion-symmetric materials offers the best nonlinear responses. A promising approach is to select materials with high dielectric contrast~\cite{heinz1991second}. However, experimental values for surface second-order susceptibility of materials and interfaces are scarce~\cite{rodriguez2008calibration, wang2010multipolar}. 

Surface nonlinearity arises from three main physical origins. First, the electric dipole contribution due to symmetry breaking at interfaces. Second, a field discontinuity at the interface, where the normal component of the electric field ($E_z$) changes rapidly, contributes to the surface nonlinearity via the electric quadrupole (nonlocal) term. Third, structural disparities between materials, such as liquid/solid interfaces, contribute to bulk-like surface nonlinearity even when dielectric constants are matched. The second and third contributions are known as electric quadrupole or nonlocal contributions to surface nonlinear susceptibility~\cite{shen1989optical, multipolar1, multipolar2}.

In this work, we present experimental investigations of ABC heterostructures in the form of SiO$_2$/TiO$_2$/Al$_2$O$_3$, deposited by atomic layer deposition. We systematically vary the thickness of each constituent material to evaluate its impact on the efficiency of second-harmonic generation. The structural properties are studied by X-ray reflectivity, scanning transmission electron microscope, and time-of-flight secondary ion mass spectrometry. The film thickness and linear optical properties are determined by spectroscopic ellipsometry. Detailed study of second harmonic generation is carried out to evaluate the second-order nonlinear susceptibility for each morphological composition.

\section{Deposition and Characterization}

\subsection*{Atomic layer deposition}
Plasma-enhanced atomic layer deposition (PEALD) was performed using a SILAYO-ICP330 system (Sentech Instruments GmbH) to fabricate \text{SiO$_2$}/\text{TiO$_2$}/\text{Al$_2$O$_3$} heterostructures. The deposition utilized oxygen plasma as a co-reactant with the following precursors: bis(diethylamino)silane (BDEAS) for \text{SiO$_2$}, tetrakis(dimethylamino)titanium (TDMAT) for \text{TiO$_2$}, and trimethylaluminum (TMA) for \text{Al$_2$O$_3$}. The inductively coupled plasma (ICP) source was operated at $100~\text{W}$ for all processes. The reactor chamber was maintained at a temperature of $100~^{\circ}\text{C}$. A 15-minute stabilization period was implemented after sample loading prior to deposition to ensure thermal stability and uniformity. Nitrogen (\text{N$_2$}) served as the precursor carrier and purge gas. The growth per cycle (GPC) is determined for single-layer depositions of approximately 20-50~nm. The deposition parameters are presented in Table~\ref{table:ALD}.

\begin{table}[htbp]
\centering
\renewcommand{\heavyrulewidth}{0.1em} 
\renewcommand{\lightrulewidth}{0.05em} 
\caption{PEALD process parameters for SiO$_2$/TiO$_2$/Al$_2$O$_3$ heterostructures.}
\scriptsize 
\renewcommand{\arraystretch}{0.8} 
\setlength{\aboverulesep}{0.2ex} 
\setlength{\belowrulesep}{0.2ex} 
\begin{tabular}{@{\hspace{2pt}}lccc@{\hspace{2pt}}}
\toprule
Material & SiO$_2$ & TiO$_2$ & Al$_2$O$_3$ \\
\midrule
Refractive index (1032 nm) & 1.45 & 2.29 & 1.63 \\
Precursor & BDEAS & TDMAT & TMA \\
\quad Pulse (ms) & 300 & 3130 & 80 \\
\quad Purge (ms) & 5000 & 8000 & 2000 \\
Plasma Gas & O$_2$ & O$_2$ & O$_2$ \\
\quad Pulse (ms) & 3000 & 5000 & 3000 \\
\quad Purge (ms) & 2000 & 5000 & 2000 \\
\quad ICP power (W) & 100 & 100 & 100 \\
\quad O$_2$ flow (sccm) & 200 & 200 & 100 \\
N$_2$ flow (sccm) & 30 & 160 & 80 \\
Sample Temperature (\si{\celsius}) & 100 & 100 & 100 \\
Growth per cycle (\si{\angstrom}/cycle) & 1.2 & 0.7 & 1.5 \\
\bottomrule
\end{tabular}
\smallskip

\footnotesize{sccm: standard cubic centimeter per minute.}
\label{table:ALD}
\end{table}

\subsection*{X-ray reflectivity}
To investigate the layered composition, layer separation, total thickness $d$, period thickness $t_{\textup{ABC}}$, and interface roughness of ABC heterostructures on a fused silica substrate, X-ray reflectivity (XRR) analysis was utilized. A Bruker AXS instrument, using a monochromatic X-ray beam (Cu-K$\upalpha$, $\lambda = 0.154\ \text{nm}$), was scanned at scattering angles from $0^\circ$ to $14^\circ$. Data analysis was performed using Bruker Leptos 7 software.
\subsection*{Transmission electron microscopy}
To examine the samples, high-resolution scanning transmission electron microscopy (HR-STEM) studies were performed using a Themis Z(3.1) 80-300 TEM instrument (Thermo Fisher Scientific) with probe Cs-aberration correction, operating at 300\nobreakspace{}nm. The microscope was equipped with a Super-X$^{\text{TM}}$ (TFS) Energy Dispersive X-ray (EDX) detector and a Gatan Image Filter (GIF) Continuum 1065ER used for Electron Energy Loss Spectroscopy (EELS). The samples were prepared by the Focused Ion Milling (FIB) lift-out technique using a Helios G4 UC FIB-SEM device operated at 30, 16 and 5\nobreakspace{}kV.

\subsection*{Secondary Ion Mass Spectrometry}
Time-of-flight secondary ion mass spectrometry (ToF-SIMS) depth profiling measurements were performed using an IONTOF TOF.SIMS\nobreakspace{}M6 instrument operating in dual-beam mode. A pulsed $\textup{Bi}^{\textup{+}}_{\textup{}}$ primary ion beam was used for analysis, while an $\textup{O}^{\textup{+}}_{\textup{2}}$ ion beam was used to sputter the sample material.
The analysis area was centered within the larger sputter crater to ensure a flat crater bottom for optimal depth resolution. The size of the sputter crater was set at $300 \times 300\,\upmu$m, and the area analyzed within this crater was measured $100 \times 100\,\upmu$m. The instrument was operated in non-interlaced mode, utilizing one sputter frame per analysis frame with a 0.1\nobreakspace{}s pause to minimize surface charging effects, which is particularly useful for insulating or semi-insulating samples. The angle of incidence for both the analysis and the sputter ion beams was 45$^\circ$ relative to the sample surface normal. A high-energy flood gun with an acceleration voltage of 300\nobreakspace{}V was also used for charge compensation. Figure \ref{fig:SIMS1} was acquired using a 1\nobreakspace{}keV sputter beam at 209\nobreakspace{}nA, whereas Figure \ref{fig:SIMS2} used 2\nobreakspace{}keV at 700\nobreakspace{}nA. The ToF-SIMS depth calibration was performed using the XRR results, under the simplifying assumption that the ablation rate remained constant throughout the entire heterostructure during the depth profile analysis.

\subsection*{Ellipsometry}
The ordinary and extraordinary refractive indices were determined for each structure using an SE850 DUV variable angle spectroscopic ellipsometer (SENTECH Instruments GmbH), and the data were evaluated in the wavelength range from 400 to $1040\ \text{nm}$ with a uniaxial Sellmeier dispersion model.

\subsection*{Nonlinear setup}
\begin{figure}[htbp]
\centering\includegraphics[width=10cm]{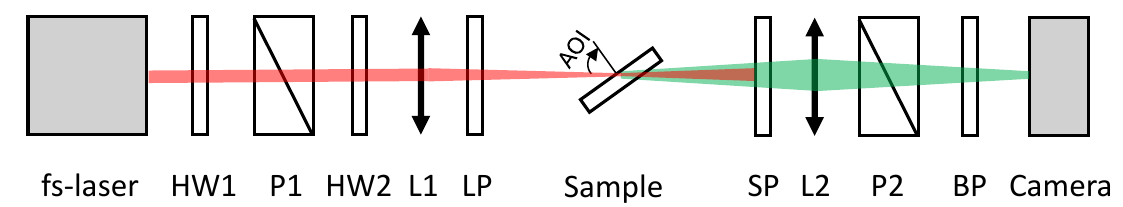}
\caption{Graphical schema of the optical setup used for second harmonic measurements of the ABC-type heterostructures. From the left: femtosecond laser, HW1: half-wave plate, P1: polarizer, HW2: half-wave plate, L1: plano-convex lens, LP: longpass filter, tilting stage with the sample, SP: shortpass filter, L2: biconvex lens, P2: Rochon prism analyzer, BP: bandpass filter, camera.}
\label{fig:setup}
\end{figure}

The following optical setup was used to investigate the nonlinear optical response of the ABC layers. A femtosecond laser (PHAROS-SP, Light Conversion) emits pulses with full-width half-maxima (FWHM) of \SI{190(10)}{\femto\second} with a central wavelength of 1032\nobreakspace{}nm and with an average power of up to \SI{1.5}{\watt} at a repetition rate of \SI{200}{\kilo\hertz}, see Fig. \ref{fig:setup}. The polarization dependence of the SHG was performed by another femtosecond laser (Satsuma, Amplitude) with a full-width half maxima of \SI{270}{\femto\second}, wavelength of 1031.4\nobreakspace{}nm and repetition rate of \SI{1}{\mega\hertz}. Power and polarization were controlled using a half-wave plate HW1 (WPH05M-1030, Thorlabs), polarizer P1 (GT10-B, Thorlabs) and a half-wave plate HW2 (Thorlabs AHWP05M-950). The laser beam was focused on the sample using the plano-convex lens L1 (LA1172-B-ML, Thorlabs) with a focal length of 400\nobreakspace{}mm, producing a beam waist of $w_0 = 55(5)\ \upmu\text{m}$. This leads to a confocal parameter of 18\nobreakspace{}mm, that is, $2\times$Rayleigh length. The loosely focused beam allows the front and back sides of the sample to remain in focus during tilting; the sample is 1\nobreakspace{}mm thick. A longpass filter LP (FELH0900, Thorlabs) was placed behind lens L1 to remove any parasitic light, especially any SHG created by the preceding optical elements. The sample was mounted on an automated rotation stage and aligned so that the laser beam remained focused on the same spot when the sample was tilted. The optical angle of incidence changed from \SI{-70}{\degree} to \SI{+85}{\degree} in steps of \SI{0.25}{\degree}. The generated SH signal from the sample was separated from the fundamental using a shortpass filter SP (FESH0800, Thorlabs). For detecting the generated SH, the sample was imaged with a CMOS camera (CS165MU1/M, Thorlabs) using a biconvex lens with a focal length of 100 mm. Additionally, a band-pass filter (FBH520-40, Thorlabs) was used to suppress the environmental light. A Rochon prism analyzer P2 (RPM10, Thorlabs) was used to determine the direction of SHG polarization. The SH signal was focused using a biconvex lens L2 (LA1608-A, Thorlabs) with a focal length of 100\nobreakspace{}mm on the camera (CS165MU1/M, Zelux). The camera calibration factor to calculate the power of the SHG is determined separately using a beta-barium borate (BBO) crystal \cite{beer2022}.

\subsection*{Nonlinear characterization}
The standard method for evaluating nonlinear second-order susceptibility $\chi^{\textup{(2)}}$ is the Maker fringe technique \cite{Maker}. It uses the tilting of the sample and the SHG intensity is collected for each angle. However, it is limited to thicker crystals with a minimum size of tens of micrometers to a few millimeters.
This limiting factor led to the development of a new method for measuring second-order susceptibility by Hermans et al. \cite{On-determination-SHG,hermans2019,koskinen2018}. This method is well suited for evaluating the $\chi^{(2)}$ values of thin films with thicknesses even less than 100\nobreakspace{}\text{nm}, where phase mismatch effects can be neglected. Similarly to the Maker fringe technique, the sample was tilted during SHG measurements using a fixed polarization fundamental frequency, and the obtained SH curve was used to fit the nonlinear coefficients. This method does not consider individual separated layers A, B, and C and each interface as a source of second harmonic, but uses an effective medium approximation. Neglecting its birefringent properties, the ABC heterostructure was represented by an average refractive index of $\mbox{$(n_{\textup{o}}+ n_{\textup{e}})/2$}$. SHG was described in terms of the surface second-order nonlinearity $\chi^{\textup{ABC}}_{\textup{s}}$ of the total heterostructure.

Any second-order nonlinear process is described by the second-order susceptibility tensor $\chi^{\textup{(2)}}_{}$, with 27 elements. However, the number of nonzero and independent elements is much smaller due to symmetry properties. The ABC heterostructure belongs to the symmetry group $C_{\infty v}$, as do other amorphous surfaces or thin achiral films with in-plane isotropy. This means that the nonzero coefficients of the second-order susceptibilities are
$\chi^{\textup{(2)}}_{{xxz}} = \chi^{\textup{(2)}}_{{yyz}} =
\chi^{\textup{(2)}}_{{zxz}} =
\chi^{\textup{(2)}}_{{yzy}}$,
$\chi^{\textup{(2)}}_{{zxx}} = \chi^{\textup{(2)}}_{{zyy}}$ and $\chi^{\textup{(2)}}_{{zzz}}$, 
where the $z$ direction is perpendicular to the layers and the $x$ and $y$ directions are parallel to the layers.
It is also assumed that these coefficients have only real values and that the measurement takes place outside the resonance region, i.e. absorption is not present.
The general description~\cite{On-determination-SHG} of the electric field components of the SH coming from a surface with $C_{\infty v}$ symmetry is described as:
\begin{equation}
\begin{aligned}
E_{2\omega,\textup{p}} = f E_{\omega,\textup{p}}^2 + g E_{\omega,\textup{s}}^2,
\end{aligned}
\end{equation}
\begin{equation}
\begin{aligned}
E_{2\omega,\textup{s}} = h E_{\omega,\textup{p}} E_{\omega,\textup{s}},
\end{aligned}
\end{equation}
where p and s represent the p- and s-polarized components and $E_{\omega}$ is the incoming electric field amplitude of the fundamental frequency. The coefficients $f$, $g$, and $h$ represent generalized layer properties such as nonlinear susceptibility tensor components, thickness, refractive index, and angle of incidence.

\begin{figure}[htbp]
\centering
\begin{subfigure}[c]{0.45\textwidth}
    \centering
    \includegraphics[scale=0.50]{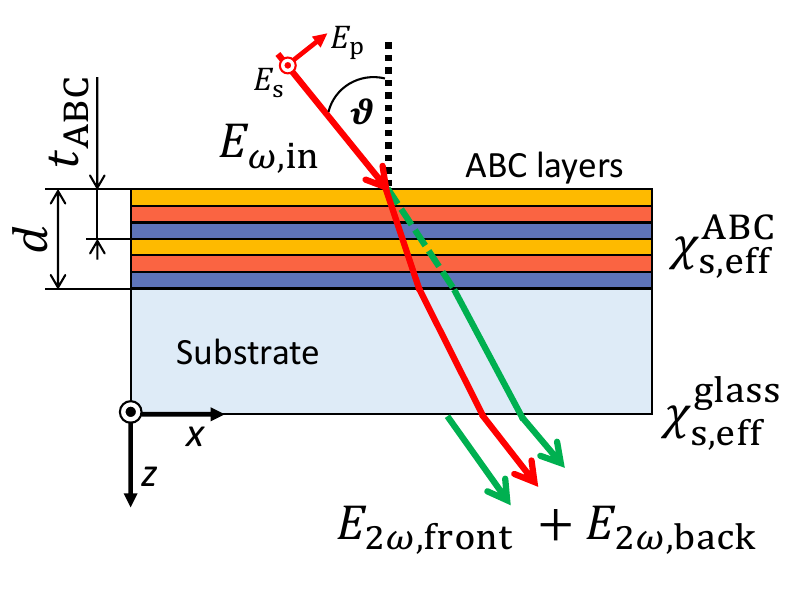}
    \caption{}
    \label{fig:picture shg abc}
\end{subfigure}
\hfill
\begin{subfigure}[c]{0.45\textwidth}
    \centering
    \includegraphics[scale=0.25]{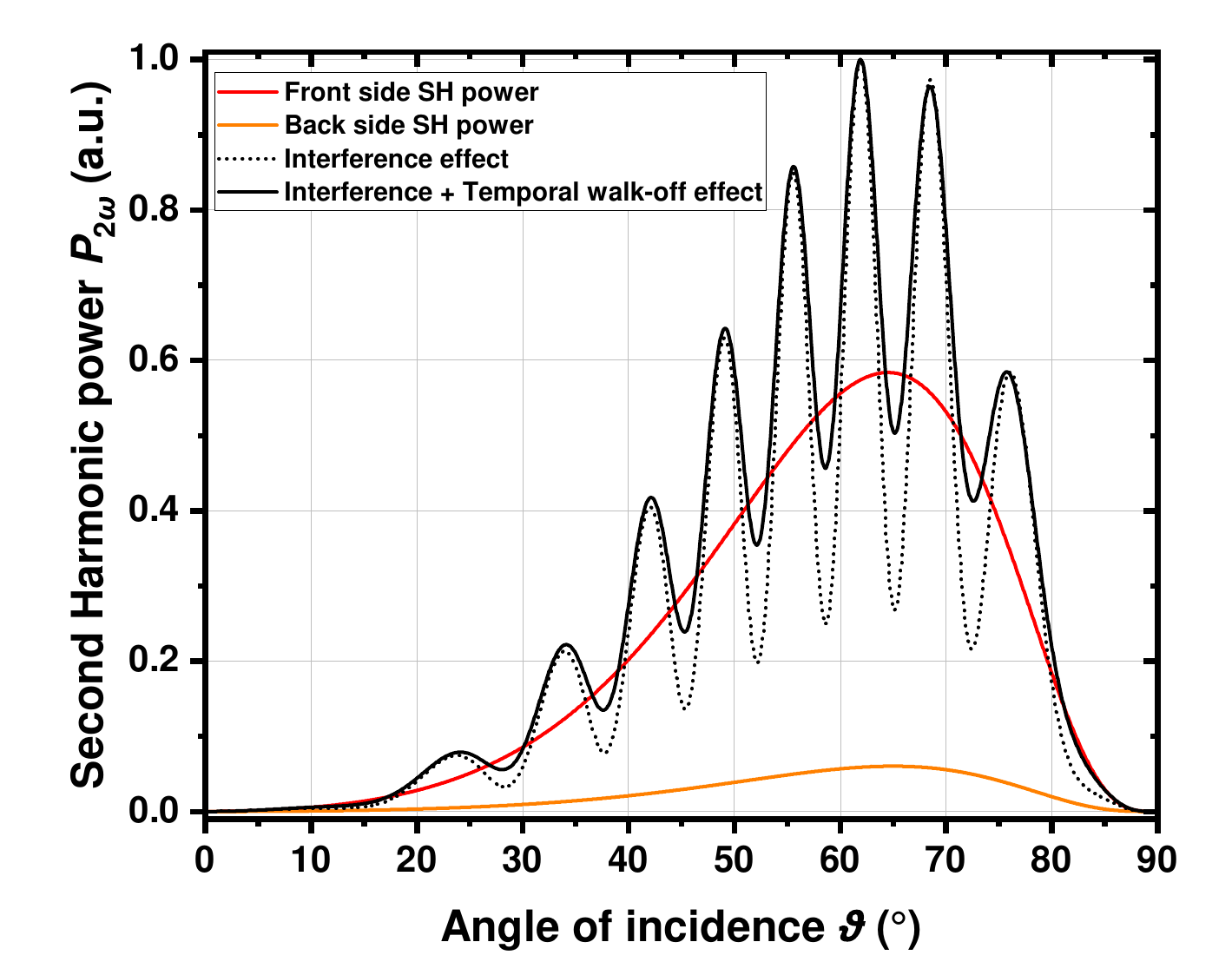}
    \caption{}
    \label{fig:STA_AOI_2}
\end{subfigure}
\caption{(a) Visualizing the angle-dependent measurement for the analysis of second harmonic generation from ABC heterostructures with the angle of incidence $\vartheta$ and the axis orientation. (b) Showing effects present in the experiment. A stronger SH signal comes from the ABC layers, and it interferes constructively and destructively with the weaker SH signal from the back side of the substrate. The temporal walk-off effect also influences the interference by making the local minima shallower.}
\label{fig:AOI}
\end{figure}

The measurement technique is based on the fact that there are two sources of SH signals, $\chi^{\textup{ABC}}_{\textup{s}}$ at the air-thin film interface at the frontside of the sample and $\chi^{\textup{glass}}_{\textup{s}}$ at the glass-air interface at the backside of the sample. Both SH signals interfere with each other, producing specific angle-dependent fringes for this method. Based on prior knowledge of the magnitude of $\chi^{\textup{glass}}_{\textup{s}}$, which is known from either the literature or from calibration measurements, the magnitude of the unknown nonlinearity $\chi^{\textup{ABC}}_{\textup{s}}$ in the thin layer can be deduced. We can obtain the bulk second-order susceptibility as $\chi^{\textup{ABC}}_{\textup{b}} = \chi^{\textup{ABC}}_{\textup{s}}/d$, where $d$ is the total thickness of the heterostructure. 
Under these approximations made by Hermans et al.\cite{On-determination-SHG,hermans2019,koskinen2018}, a monochromatic p-polarized plane wave of the fundamental frequency of the amplitude of the electric field $E_{\omega,\textup{p}}$ produces a transmitted p-polarized SH electric field $E_{2\omega,\textup{total}}$ as 
\begin{equation}
\begin{aligned}
E_{2\omega,\textup{total}} &= E_{2\omega,\textup{front}} + E_{2\omega,\textup{back}} \\
&= -j\frac{\omega}{2c} t^2_{\textup{air,ABC}} E^2_{\omega,\textup{in}} \Bigg[ \frac{T_{\textup{ABC,glass}} T_{\textup{glass,air}}}{N_{\textup{ABC}} \cos(\Theta_{\textup{ABC}})} \chi^{\textup{ABC}}_{\textup{s,eff}}\exp\left(-j \frac{2\omega N_{\textup{glass}} \cos(\Theta_{\textup{glass}}) L_{\textup{glass}}}{c}\right) \\
&\quad - \frac{t^2_{\textup{ABC,glass}} T_{\textup{glass,air}} }{N_{\textup{glass}} \cos(\Theta_{\textup{glass}}) } \chi^{\textup{glass}}_{\textup{s,eff}}\exp\left(-j \frac{2\omega n_{\textup{glass}} \cos(\theta_{\textup{glass}}) L_{\textup{glass}}}{c}\right) \Bigg],
\end{aligned}
\label{E_total}
\end{equation}
where $\chi^{{i}}_{\textup{s,eff}}$ is the effective surface second-order susceptibility 
\begin{equation}
\begin{aligned}
\chi_{\textup{s,eff}}^i &= \chi_{\textup{s},xxz}^i \sin(2\theta_i) \cos(\Theta_i) + \chi_{\textup{s},zxx}^i \sin(\Theta_i) \cos^2(\theta_i) + \chi_{\textup{s},zzz}^i \sin^2(\theta_i) \sin(\Theta_i) \\
              &\approx \left( \chi_{\textup{s},zxx}^i + 2 \chi_{\textup{s},xxz}^i \right) \sin(\Theta_i) \cos^2(\theta_i) + \chi_{\textup{s},zzz}^i \sin^2(\theta_i) \sin(\Theta_i).
\end{aligned}
\label{chi}
\end{equation}
The parameters for the fundamental frequency (FF) $\omega$ are denoted by lower case letters, and for the SH by upper case. The parameters $T_{i,j}$ and $t_{i,j}$ are the Fresnel transmission coefficients for p-polarized light that transmits from medium $i$ to medium $j$. The variables $N_i$ and $n_i$ are refractive indices of medium $i$. The variables $\Theta_i$ and $\theta_i$ are angles of propagation in each medium $i$ with respect to the surface normal $z$. The speed of light is $c$ and the thickness of the substrate is $L_{\textup{glass}}$. The refractive indices for the glass substrate made of pure fused silica are $N_{\textup{glass}} = 1.4615$ and $n_{\textup{glass}} = 1.4500$. The approximation in Eq. \ref{chi} is valid only for materials with low dispersion with $\Theta_i\approx\theta_i$. Due to this approximation, the values $\chi_{\textup{s},zxx}^{\textup{ABC}}$ and $\chi_{\textup{s},xxz}^{\textup{ABC}}$ cannot be fitted independently, but only as a sum $A_{\textup{s},zx}^{\textup{ABC}} = \chi_{\textup{s},zxx}^{\textup{ABC}} + 2\chi_{\textup{s},xxz}^{\textup{ABC}}$. For substrate nonlinearity we used literature values \cite{rodriguez2008calibration} at the wavelength 1064\nobreakspace{}nm: $\chi_{\textup{s},zxx}^{\textup{glass}} = 3.8 \times10^{-22}$\nobreakspace{}pm/V, $\chi_{\textup{s},xxz}^{\textup{glass}} = 7.9\times10^{-22}$\nobreakspace{}pm/V, $\chi_{\textup{s},zzz}^{\textup{glass}} = 5.9\times10^{-22}$\nobreakspace{}pm/V and we apply the Miller's rule
\begin{equation}
\begin{aligned}
\frac{\chi^{(2)}(2\omega)}{\chi^{(1)}(2\omega) \cdot \left|\chi^{(1)}(\omega)\right|^2} = \text{const.}
\end{aligned},
\end{equation}
with $\chi^{(1)}$ as the linear susceptibility of the material, to estimate the values at a wavelength of 1032\nobreakspace{}nm. Because a femtosecond laser is used, the group velocity mismatch between the fundamental and its SH introduces a temporal delay between the SH generated at the front and back surfaces of the sample. By the time the SH generated at the front surface reaches the back surface, an additional SH contribution has already formed there. This relative delay between the two SH contributions is known as the temporal walk-off time $t_{\text{walk-off}}$. This causes destructive interference to be incomplete, and thus, the contrast of the interference fringes is decreased. When the temporal walk-off is not taken into account, the value of $\chi_{\textup{}}^{(2)}$ is overestimated. For our fused silica sample, 1\nobreakspace{}mm thick, the value was measured to be $t_{\text{walk-off}} = 94$\nobreakspace{}fs. This value corresponds also well with the calculation of the delay based on the group velocity of each pulse in the substrate. Therefore, the model was modified by Hermans et al.\cite{On-determination-SHG,hermans2019,koskinen2018} for the pulsed laser with the temporal walk-off effect as 
\begin{equation}
\begin{aligned}
P_{2\omega} = K_2 \int_{-\infty}^{+\infty} 
\left[ E_{2\omega, \text{front}} \ \text{sech}^2 \left(\frac{t}{\frac{\Delta t }{2 \ln \left( 1 + \sqrt{2} \right) }}\right) + E_{2\omega, \text{back}} \ \text{sech}^2 \left(\frac{t + t_{\text{walk-off}}}{\frac{\Delta t }{2 \ln \left( 1 + \sqrt{2} \right) }}\right) \right]^2 \, dt
\end{aligned},
\label{K2}
\end{equation}
where $K_2$ is the calibration constant, $\Delta t$ is the pulse duration at FWHM. However, the walk-off time is not constant; it changes as the angle of incidence increases, and the optical path through the substrate increases as well. Then, the relation is $t_{\text{walk-off}} = {t_{\text{walk-off,0}}}/{\cos(\theta_{\text{glass}})}$, where $t_{\text{walk-off,0}}$ is walk-off time at the normal angle of incidence. The walk-off effect is also visualized in Fig. \ref{fig:AOI}.
By substituting Eq. \ref{K2} with Eq. \ref{E_total}, we can obtain the desired second-order susceptibility together with the other fitting parameters $\chi_{\textup{s},zzz}^{\textup{ABC}}$, $A_{\textup{s},zx}^{\textup{ABC}}$, $K_2$, and $L_{\textup{glass}}$. The thickness of the glass substrate must also be fitted for each measurement, as the position of the fringes in the second harmonic is highly sensitive to variations in substrate thickness. Dividing the surface susceptibility by the total thickness of the heterostructure $\chi_{\textup{bulk}}^{\textup{ABC}}=\chi_{\textup{s}}^{\textup{ABC}}/d$, the bulk second-order susceptibility is determined. This simplified version of the model neglects multiple reflections in the film and substrate while maintaining validity, as previously tested. Therefore, we keep the angle of incidence below $60^\circ$ for the analysis. For additional details, see Supplement 1.

The primary source of uncertainty in the $\chi^{(2)}$ evaluation stems from the uncertainty in determining the angle of incidence, estimated as a systematic error of $\pm 1^\circ$. This leads to an uncertainty in the $\chi^{(2)}$ value that is substantially greater than the standard deviations derived from the nonlinear regression analysis. We estimate the uncertainty to be approximately 10\% for the stronger SH signal samples and up to 20\% for the weaker ones.

The method currently has two limitations. First, it assumes no phase mismatch in the second harmonic, rendering it applicable only to thin films. Second, the SH signal originates not only from the ABC stack but also from the interfaces between the first deposited layer of material A and the substrate, and between the last deposited layer of material C and air. The method fails to account for these additional sources of second harmonic generation, which can lead to an overestimation of the second-order nonlinear susceptibility of the ABC heterostructure, particularly in samples with a limited number of ABC interfaces where the C-air interface may become significant. To estimate this effect, single layers of each material were also investigated. 

\section{Results and discussion}

Atomic layer deposition (ALD) enables precise control over thin-film thickness via self-limited growth, with growth rates per cycle (GPC) of $0.12$\nobreakspace{}nn/cycle for SiO$_2$ (A), $0.07$\nobreakspace{}nn/cycle for TiO$_2$ (B), and $0.15$\nobreakspace{}nn/cycle for Al$_2$O$_3$ (C). To investigate the nonlinear interface response of ABC heterostructures, multiple samples with a constant total deposition thickness of approximately 60\nobreakspace{}nm were fabricated, while systematically varying the ABC period thickness $t_{\text{ABC}}$ from 60\nobreakspace{}nm down to the atomic scale of 0.3\nobreakspace{}nm and it is easily achieved by ALD by controlling the number of cycles. This approach enabled the study of nonlinear responses by altering the density of interfaces per unit thickness and exploring different morphologies across the samples, with the goal of understanding how these variations influence the properties of heterostructures. Each oxide material is represented approximately equally in the period $t_{\text{ABC}}$. The ABC heterostructures consist of around 300--380 atomic layers for a 60\nobreakspace{}nm thick period and only about two atomic layers for an ABC period thickness of 0.3\nobreakspace{}nm, assuming a Si--O bond length of 0.15\nobreakspace{}nm, a Ti--O bond length of 0.20\nobreakspace{}nm, and an Al--O bond length of 0.16\nobreakspace{}nm \cite{Si-O,Ti-O,Al-O}, suggesting ultimate physical limits for ABC layers.

\begin{figure}[htbp]
  \centering
  
  \begin{subfigure}[t]{0.35\textwidth}
    \centering
    \includegraphics[width=\textwidth]{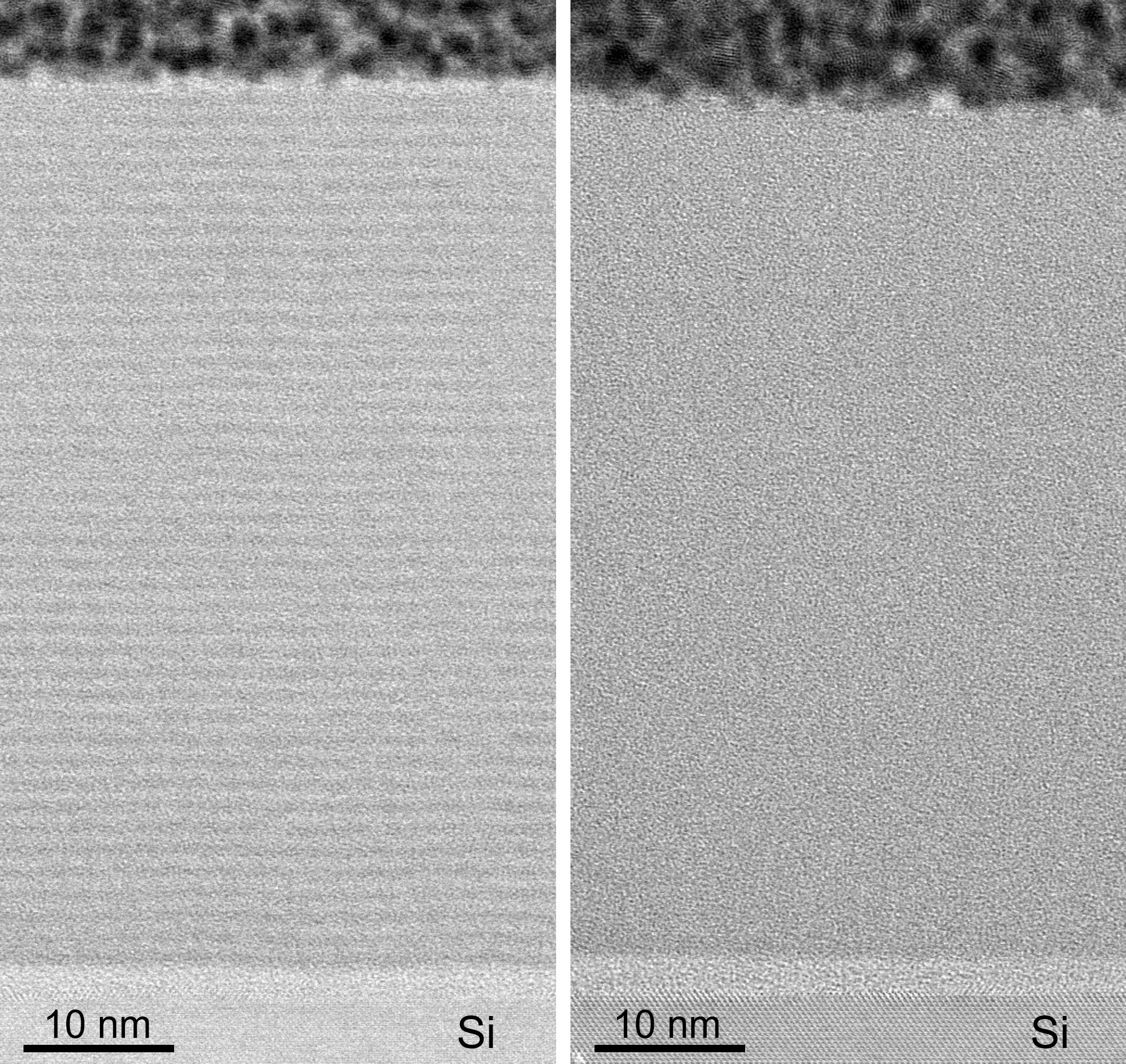}

  \end{subfigure}

  \caption{ADF STEM images showing the \text{SiO$_2$/TiO$_2$/Al$_2$O$_3$} nanolaminates on the silicon wafer substrate of: (left) sample S3 with the ABC period thickness $t_{\textup{ABC}}$ of 1.5\nobreakspace{}nm; TiO$_{2}$ layers appear darker; (right) sample S2 with $t_{\textup{ABC}}$ of 0.7\nobreakspace{}nm. The layers are no longer visible. The dark area on the top of both nanolaminates is the Pt/C protection layer used for TEM specimen preparation.}
  \label{fig:TEM}
\end{figure}

Scanning transmission electron microscope (STEM) images were collected using a high-angle annular dark-field (HAADF) and annular dark field (ADF) detectors to reveal the structural characteristics of the \text{SiO$_2$/TiO$_2$/Al$_2$O$_3$} nanolaminates on a silicon wafer substrate in selected samples S3 and S2. The ADF images are shown in the Fig. \ref{fig:TEM}. Sample S3 exhibits a clear ABC period thickness $t_{\text{ABC}}$ of 1.5\nobreakspace{}nm, with clearly visible distinct layers. On the other hand, sample S2, with a reduced $t_{\text{ABC}}$ of 0.7\nobreakspace{}nm, shows no recognizable layer separation, indicating a loss of structural distinction on this scale.

\begin{figure}[htbp]
\centering
\includegraphics[scale=0.25]{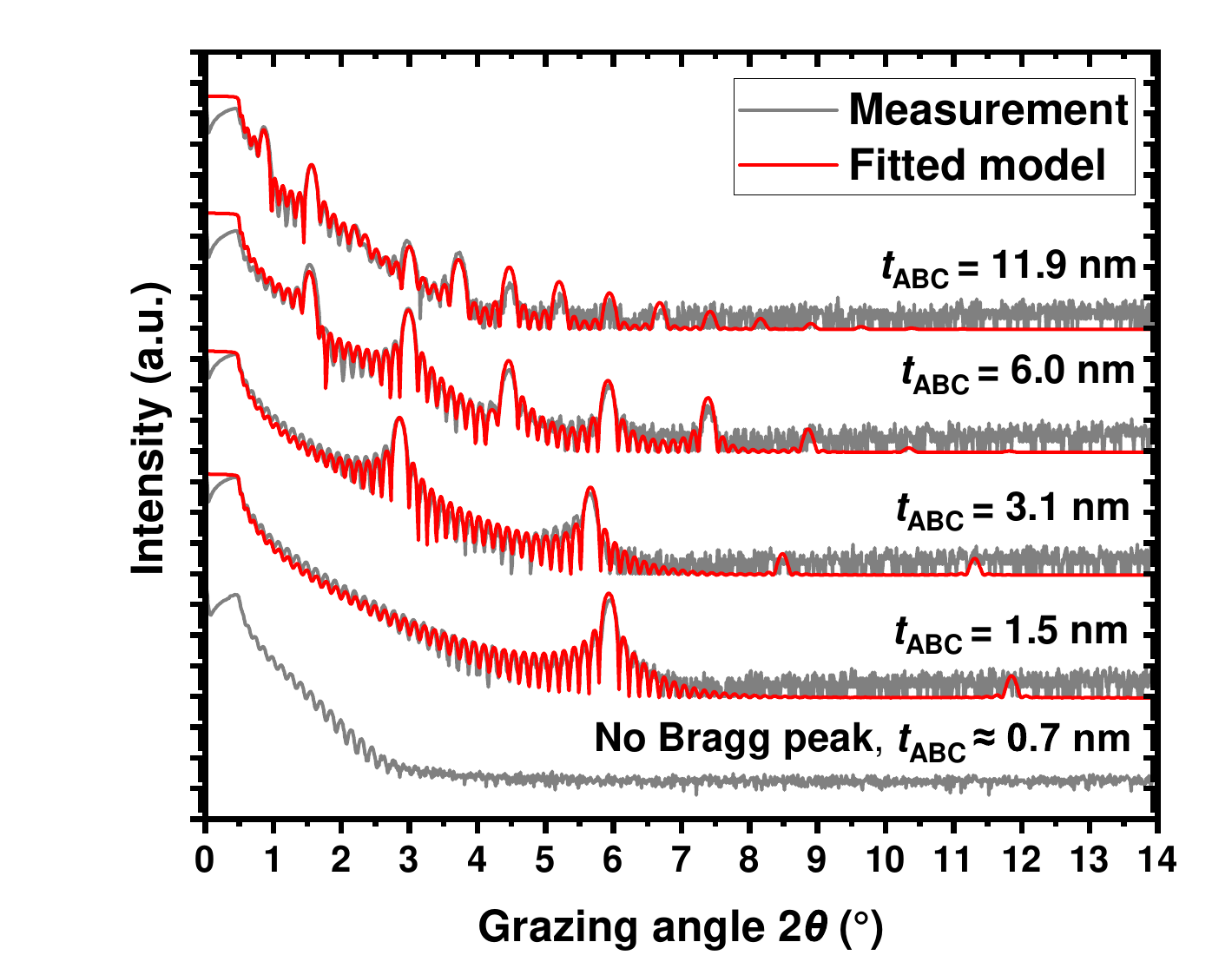}
\caption{Results of the X-ray reflectivity measurements proving the existence of the layers and verifying
the period thickness $t_{\text{ABC}}$. Sample ID from the top: S13, S11, S6, S3, S2.}
\label{fig:XRR}
\end{figure}

 To verify the precise thickness of the layers in each composition, X-ray reflectivity (XRR) was employed. The measurement spectra of the heterostructures are presented in Fig. \ref{fig:XRR}. The Bragg peaks are clearly visible for all samples except for the heterostructure with a period of 0.7 nm (sample S2), suggesting notable layer intermixing and a lack of clearly separated layers. As the period of the heterostructure decreases, the first Bragg peak shifts to the higher grazing angle. The XRR analysis shows the existence of separated nanolaminates and the excellent agreement with the targeted ABC period thickness. The analysis also provided the total deposition thickness and provided the average interface roughness of all three types of interfaces in the heterostructure, see Table \ref{table:XRR}.

\begin{table}[htbp]
\centering
\renewcommand{\heavyrulewidth}{0.1em} 
\renewcommand{\lightrulewidth}{0.05em} 
\setlength{\tabcolsep}{2pt} 
\caption{Results of the X-ray reflectivity measurements for selected SiO$_2$/TiO$_2$/Al$_2$O$_3$ heterostructures.}
\scriptsize 
\renewcommand{\arraystretch}{0.8} 
\setlength{\aboverulesep}{0.2ex} 
\setlength{\belowrulesep}{0.2ex} 
\begin{tabular}{@{\hspace{2pt}}lccccccc@{\hspace{2pt}}}
\toprule
\parbox[c]{1.2cm}{\centering Sample ID} 
& \parbox[c]{1.8cm}{\centering Total thickness $d$ (nm)}
& \parbox[c]{1.8cm}{\centering Period thickness $t_{\textup{ABC}}$ (nm)} 
& \parbox[c]{1.8cm}{\centering Layer thickness SiO$_2$ (nm)} 
& \parbox[c]{1.8cm}{\centering Layer thickness TiO$_2$ (nm)} 
& \parbox[c]{1.8cm}{\centering Layer thickness Al$_2$O$_3$ (nm)} 
& \parbox[c]{1.8cm}{\centering Average interface roughness (nm)} \\
\toprule
S3 & 59.7 & 1.5 & 0.4 & 0.4 & 0.7 & 0.4 \\
\midrule
S6 & 62.6 & 3.1 & 1.4 & 1.0 & 0.8 & 0.3 \\
\midrule
S11 & 59.9 & 6.0 & 1.5 & 1.7 & 2.8 & 0.4 \\
\midrule
S13 & 59.6 & 11.9 & 3.4 & 2.8 & 5.7 & 0.7 \\
\bottomrule
\end{tabular}
\label{table:XRR}
\end{table}

\begin{figure}[htbp]
\centering
\begin{subfigure}[t]{0.49\textwidth}
    \centering
    \includegraphics[scale=0.25]{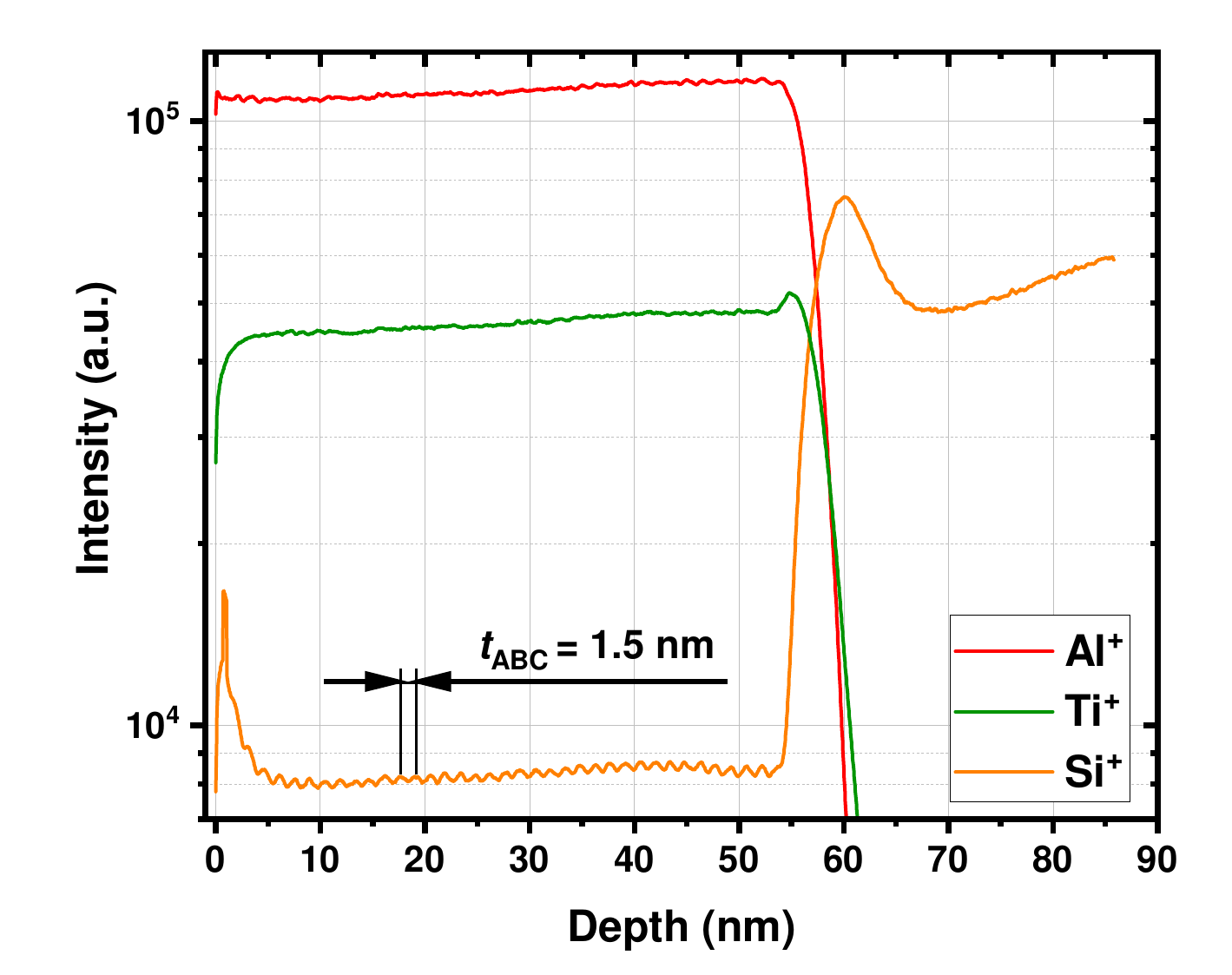}
    \caption{}
   \label{fig:SIMS1}
\end{subfigure}
\hfill
\begin{subfigure}[t]{0.49\textwidth}
    \centering
    \includegraphics[scale=0.25]{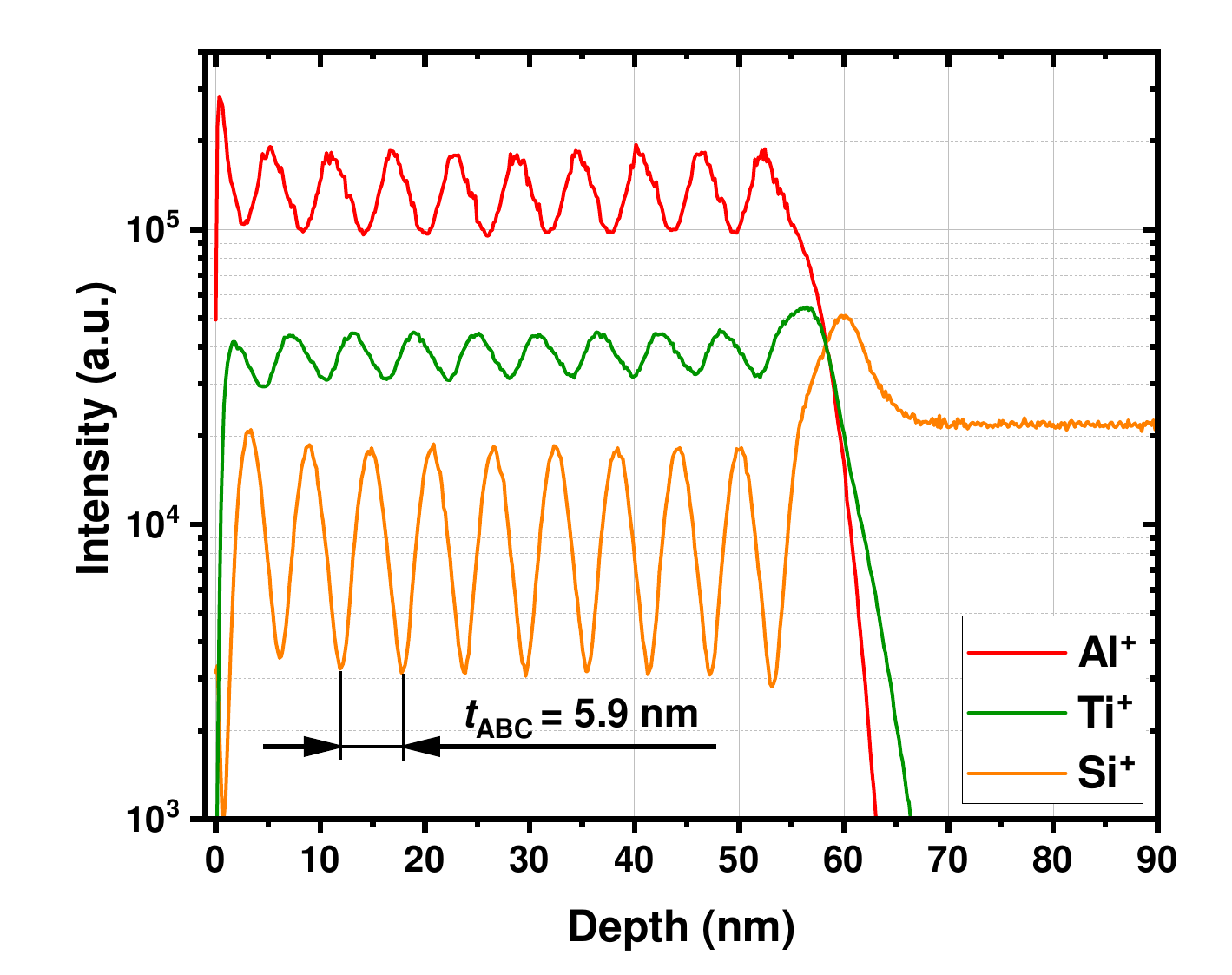}
    \caption{}
    \label{fig:SIMS2}
\end{subfigure}
\caption{Depth profiles of positive secondary ion yields measured by TOF-SIMS in the ABC heterostructure: (a) sample S3 and (b) sample S11.}
\label{fig:SIMS}
\end{figure}

Time-of-flight secondary ion mass spectrometry (TOF-SIMS) depth profiles of the heterostructures are shown in Fig.~\ref{fig:SIMS}. 
In sample S3 ($t_{\text{ABC}} = 1.5$\,nm), periodic modulation is observed only for Si$^{+}$ ions, indicating distinct SiO$_{2}$ layers, while the Al$^{+}$ and Ti$^{+}$ signals do not show a clear periodicity. 
In contrast, sample S11 ($t_{\text{ABC}} = 5.9$\,nm) exhibits periodic modulation of all constituent cations (Si$^{+}$, Al$^{+}$, Ti$^{+}$). 
The lack of resolved Al$_{2}$O$_{3}$ and TiO$_{2}$ layering in S3 does not rule out separated ABC layers, as it may arise from limited depth resolution due to differential sputtering yields, intermixing, or sample-preparation effects. 
Independent X-ray reflectivity (XRR) and scanning transmission electron microscopy (STEM) analyzes confirm the presence of separated layers in S3 with the 1.5\nobreakspace{}nm period.

\begin{table}[htbp]
\centering
\renewcommand{\heavyrulewidth}{0.1em} 
\renewcommand{\lightrulewidth}{0.05em} 
\setlength{\tabcolsep}{2pt} 
\caption{Sample parameters for SiO$_2$/TiO$_2$/Al$_2$O$_3$ heterostructures and their optical properties.}
\scriptsize 
\renewcommand{\arraystretch}{0.8} 
\setlength{\aboverulesep}{0.2ex} 
\setlength{\belowrulesep}{0.2ex} 
\begin{tabular}{@{\hspace{2pt}}lcccccccccccc@{\hspace{2pt}}}
\toprule
\parbox[c]{0.8cm}{\centering Sample ID} 
& \parbox[c]{1.8cm}{\centering Deposition cycles SiO$_2$/TiO$_2$/Al$_2$O$_3$} 
& \parbox[c]{0.8cm}{\centering Super cycles $M$} 
& \parbox[c]{2.0cm}{\centering Layer thickness (nm) SiO$_2$/TiO$_2$/Al$_2$O$_3$} 
& \parbox[c]{1.2cm}{\centering Period thickness $t_{\textup{ABC}}$ (nm)} 
& \parbox[c]{1.0cm}{\centering Total thickness $d$ (nm)} 
& \parbox[c]{0.8cm}{\centering $n_\text{o}$ (1032 nm)} 
& \parbox[c]{0.8cm}{\centering $n_\text{e}$ (1032 nm)} 
& \parbox[c]{0.8cm}{\centering $N_\text{o}$ (516 nm)} 
& \parbox[c]{0.8cm}{\centering $N_\text{e}$ (516 nm)} 
& \parbox[c]{0.9cm}{\centering $A_\text{zx}$ (pm/V)} 
& \parbox[c]{0.9cm}{\centering $\chi_\text{zzz}$ (pm/V)} \\
\toprule
\rowcolor{lightgrey} S1* & 1:1:1 & 561 & 0.12/0.07/0.15 & 0.3 & 57 & 1.76 & - & 1.80 & - & 0.29 & 0.54 \\
\midrule
\rowcolor{lightgrey} S2* & 2:3:2 & 80 & 0.24/0.21/0.3 & 0.7 & 57 & 1.81 & - & 1.86 & - & 0.50 & 0.96 \\
\midrule
\rowcolor{lightgrey} S3 & 4:6:4 & 40 & 0.48/0.42/0.6 & 1.5 & 59 & 1.77 & 1.75 & 1.83 & 1.81 & 0.80 & 2.0 \\
\midrule
S4 & 4:6:4 & 120 & 0.48/0.42/0.6 & 1.5 & 174 & 1.79 & 1.75 & 1.84 & 1.80 & 0.68 & 2.1 \\
\midrule
S5 & 4:6:4 & 160 & 0.48/0.42/0.6 & 1.5 & 233 & 1.78 & 1.76 & 1.84 & 1.80 & 0.77 & 2.5 \\
\midrule
\rowcolor{lightgrey} S6 & 8:12:8 & 20 & 0.96/0.84/1.2 & 3.0 & 63 & 1.79 & 1.71 & 1.85 & 1.75 & 0.61 & 1.8 \\
\midrule
S7 & 8:12:8 & 31 & 0.96/0.84/1.2 & 3.0 & 93 & 1.79 & 1.69 & 1.85 & 1.73 & 0.52 & 1.6 \\
\midrule
S8 & 8:12:8 & 39 & 0.96/0.84/1.2 & 3.0 & 119 & 1.79 & 1.69 & 1.85 & 1.73 & 0.55 & 1.6 \\
\midrule
S9 & 8:12:8 & 40 & 0.96/0.84/1.2 & 3.0 & 116 & 1.78 & 1.70 & 1.84 & 1.74 & 0.43 & 1.4 \\
\midrule
S10 & 8:12:8 & 70 & 0.96/0.84/1.2 & 3.0 & 215 & 1.78 & 1.71 & 1.84 & 1.75 & 0.69 & 1.9 \\
\midrule
\rowcolor{lightgrey} S11 & 16:24:16 & 10 & 1.92/1.68/2.4 & 6.0 & 60 & 1.77 & 1.66 & 1.83 & 1.71 & 0.15 & 0.86 \\
\midrule
S12 & 16:24:16 & 20 & 1.92/1.68/2.4 & 6.0 & 114 & 1.79 & 1.64 & 1.84 & 1.67 & 0.070 & 0.96 \\
\midrule
\rowcolor{lightgrey} S13 & 32:48:32 & 5 & 3.84/3.36/4.8 & 12 & 60 & 1.78 & 1.64 & 1.84 & 1.68 & 0.090 & 0.70 \\
\midrule
\rowcolor{lightgrey} S14 & 80:120:80 & 2 & 9.6/8.4/12 & 30 & 59 & 1.82 & 1.68 & 1.88 & 1.71 & 0.092 & 0.52 \\
\midrule
\rowcolor{lightgrey} S15$\dagger$ & 160:240:160 & 1 & 19.2/16.8/24 & 60 & 60 & - & - & - & - & - & - \\
\bottomrule
\end{tabular}

\footnotesize{
*: No birefringence was observed. $\dagger$: The effective medium approximation is no longer valid in the wavelength range from 400\,nm to 1040\,nm based on the ellipsometric evaluation. Layer and period thickness $t_{\textup{ABC}}$ is calculated based on the growth rate per cycle, and the total thickness $d$ is based on ellipsometric evaluation. The uncertainty for the refractive index is $\pm 0.01$. Samples with targeted 60\,nm of the total thickness are highlighted in gray.}
\label{table:big}
\end{table}

\begin{figure}[htbp]
\centering
\includegraphics[scale=0.25]{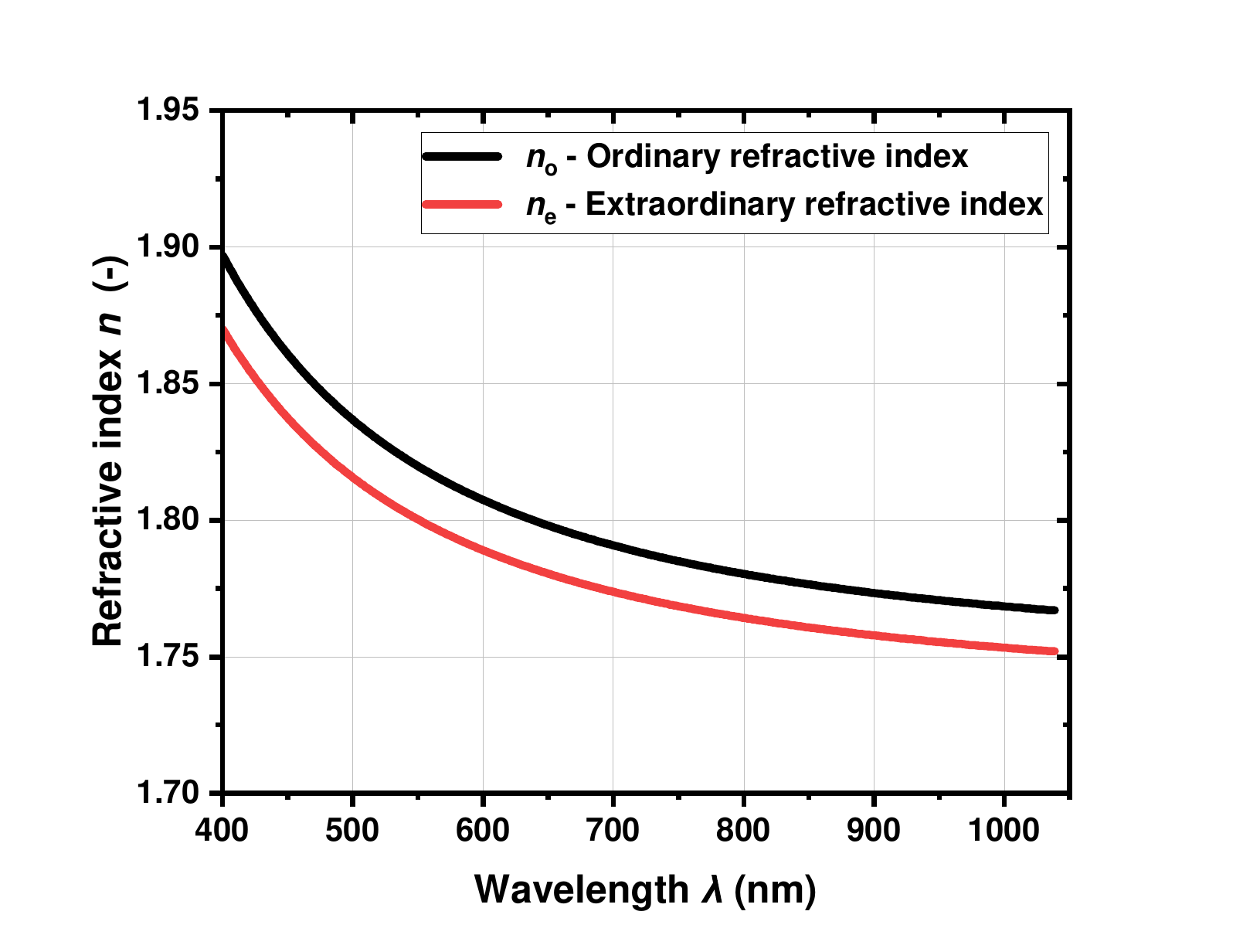}
\caption{Ordinary and extraordinary refractive indices of \text{SiO$_2$/TiO$_2$/Al$_2$O$_3$} nanolaminates of the sample S3 show the presence of the form birefringence.}
\label{fig:none sta07}
\end{figure}

ABC-type heterostructures have birefringent properties \cite{hermans2019}. For monocrystals, birefringence can be explained by the anisotropic electrical properties of the molecules that make up the crystals. However, birefringence can also be caused by anisotropy if the material is arranged on an order of magnitude larger than the size of the molecules and this distance is still smaller than the wavelength of light. The ABC-type heterostructure, composed of three materials arranged in thin layers, acts as an effective medium exhibiting form birefringence and functioning as a negative uniaxial crystal. At the same time, the entire structure is amorphous, and the optical axis is perpendicular to the layers \cite{born2013principles, Formbiro}. Ellipsometry confirms the presence of the negative birefringence, and the dispersion curve of the ordinary and extraordinary refractive index are presented in Fig. \ref{fig:none sta07}. The magnitude of birefringence, \(\Delta n = n_\text{e} - n_\text{o}\), depends on the thickness of the ABC period, achieving up to \(\Delta n = 0.15\) for sample S12, and completely disappears in samples with nonexistent layers S1 and S2. Optical properties are summarized in Table \ref{table:big}.

\begin{figure}[htbp]
\centering
\includegraphics[scale=0.25]{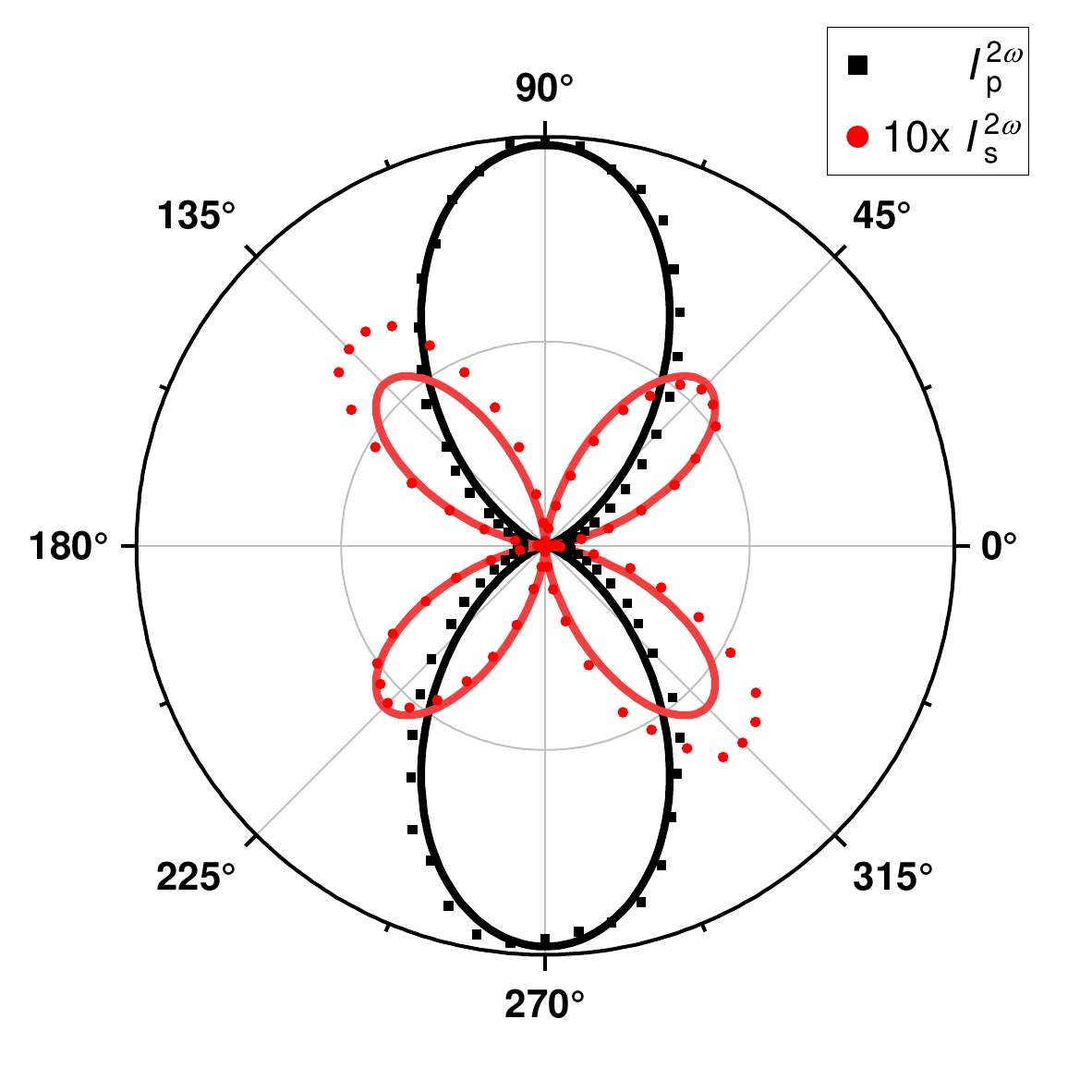}
\caption{Normalized SHG power for s-~and p-polarizations as a function of the polarization angle $\varphi$ of the fundamental corresponding with the symmetry group $C_{\infty v}$, measured for the sample~S3. Dots represent measurements and lines normalized values according to the theory.}
\label{fig:STA_polarization}
\end{figure}

Nonlinear polarization measurements were conducted to confirm the symmetry properties of ABC stacks. The linear polarization angle of the fundamental field, $\varphi$, was varied, while the second harmonic analyzer was fixed along the s- or p-polarization direction. At $\varphi=0^\circ$, the pump light is s-polarized, and at $\varphi=90^\circ$, it is p-polarized. Figure~\ref{fig:STA_polarization} displays the s- and p-polarized SH signals from sample S3 as a function of $\varphi$, with an incidence angle of $\vartheta=63^\circ$. The polarization data align with the $C_{\infty v}$ symmetry group, with the maximum SH signal observed when both the fundamental and SH fields are p-polarized. The discrepancy between the theoretical and experimental s-polarized SHG data may stem from the strain present in the ABC stack, which alters the nonlinear susceptibility tensor via the photoelastic effect \cite{Mennel2018}.

\begin{figure}[htbp]
\centering
\begin{subfigure}[t]{0.49\textwidth}
    \centering
    \includegraphics[scale=0.25]{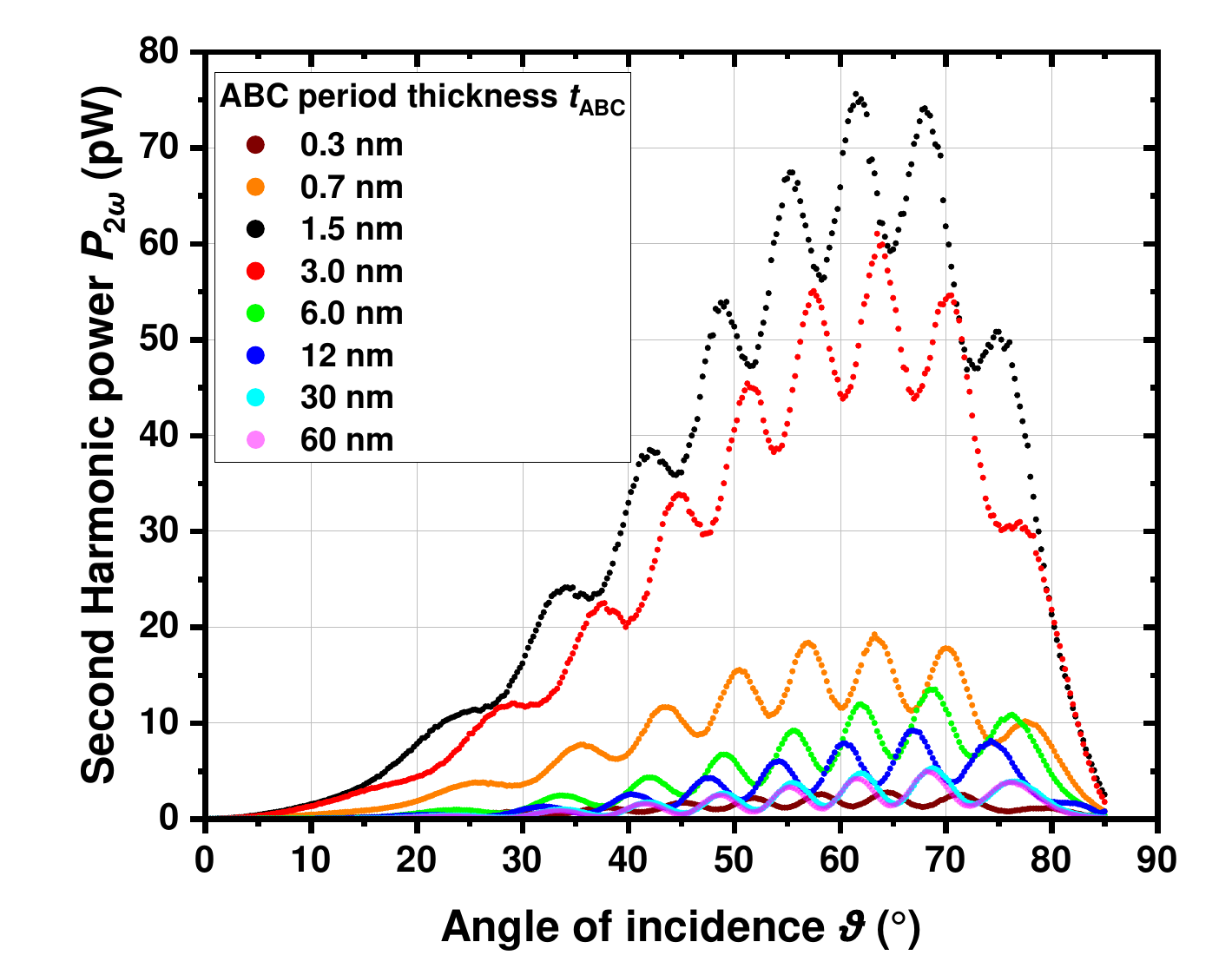}
    \caption{}
    \label{fig:STA_AOI}
\end{subfigure}
\hfill
\begin{subfigure}[t]{0.49\textwidth}
    \centering
    \includegraphics[scale=0.25]{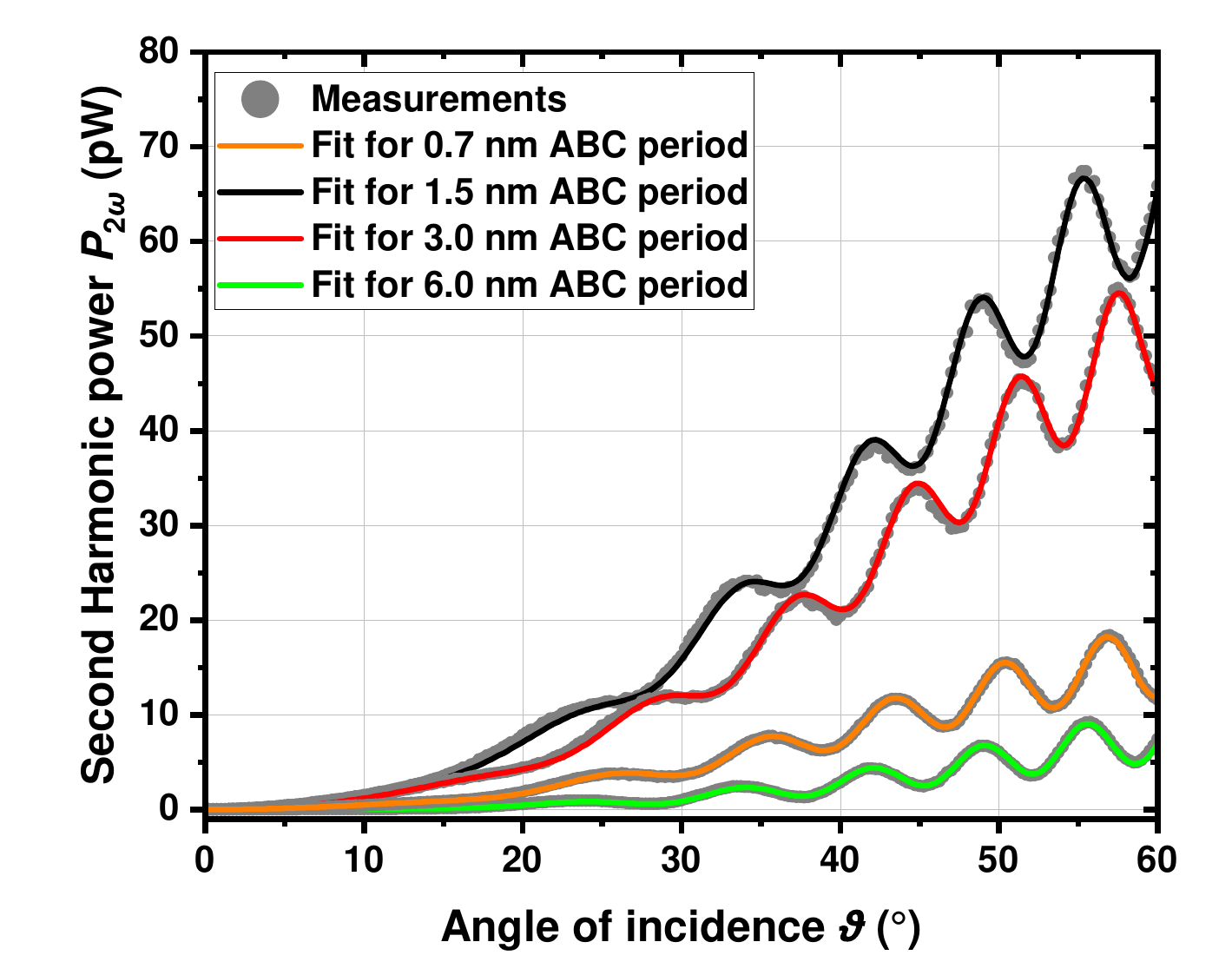}
    \caption{}
    \label{fig:STA07_AOI_fit}
\end{subfigure}
\caption{(a) Angle-dependent measurements of the second harmonic from \text{SiO$_2$/TiO$_2$/Al$_2$O$_3$} heterostructures deposited on 1\nobreakspace{}mm thick fused silica substrate. The total deposited thickness is approximately 60\nobreakspace{}nm across all samples; only the morphology, i.e., the ABC period thickness $t_{\textup{ABC}}$, varies from sample to sample. Both FF with an average pump power $P=0.5$\nobreakspace{}W and SH are p-polarized. (b) Fitting of SHG data from selected samples to analyze the second-order susceptibility of the ABC heterostructures.}
\label{fig:combined_STA}
\end{figure}

Based on the experimental results for the p-polarized fundamental and the second harmonic in Fig. \ref{fig:STA_AOI}, the intensity of the second harmonic increases with decreasing ABC period thickness from 60\nobreakspace{}nm until 1.5\nobreakspace{}nm, where a maximum is observed. For periods thinner than 1.5\nobreakspace{}nm, the second harmonic signal decreases abruptly. As each layer is approximately 0.25\nobreakspace{}nm thick, intermixing at the interfaces likely degrades layer quality, and incomplete layer closure further reduces the SH signal. This hypothesis is supported by STEM imaging and XRR analysis of sample S2, which show no clear layers for an ABC period of $t_{\textup{ABC}}=0.7$\nobreakspace{}nm. The nonlinear model for the determination of $\chi^{\textup{(2)}}_{}$ is fitted to all SHG data across varying ABC periods, with selected curves shown in Fig. \ref{fig:STA07_AOI_fit}.

\begin{figure}[htbp]
\centering
    \includegraphics[scale=0.25]{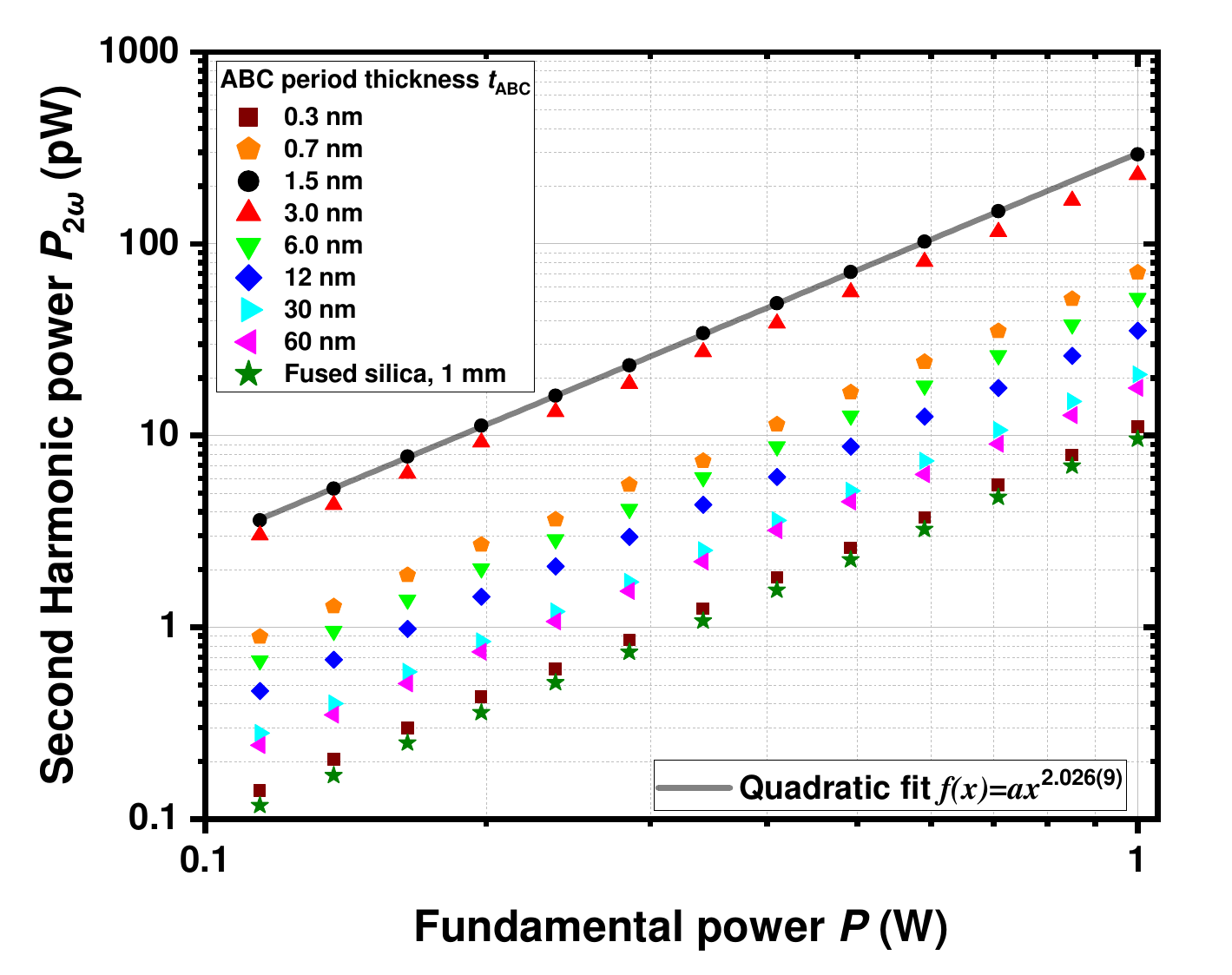}
\caption{Average second harmonic power dependence on average fundamental power in \text{SiO$_2$/TiO$_2$/Al$_2$O$_3$} heterostructures, measured at the angle of incidence of maximum second harmonic intensity for both p-polarized fundamental and second harmonic. Quadratic power dependence of the second harmonic for sample S3 with $t_{\textup{ABC}}=1.5$\nobreakspace{}nm, fitted to the data.}
\label{fig:STA_power}
\end{figure}

The same trend is also reported for power dependency measurements. The samples are aligned at the incidence angle that yields the maximum SHG signal for p-polarized fundamental light. For comparison, an uncoated substrate of fused silica with surface second-order nonlinearity is also included. Verification of second-harmonic generation is achieved by observing its quadratic power dependency, and excellent agreement is found by achieving a slope line of $2.026(9)$ for sample S3, see Fig.\nobreakspace{}\ref{fig:STA_power}.

\begin{figure}[htbp]
\centering
\includegraphics[scale=0.25]
{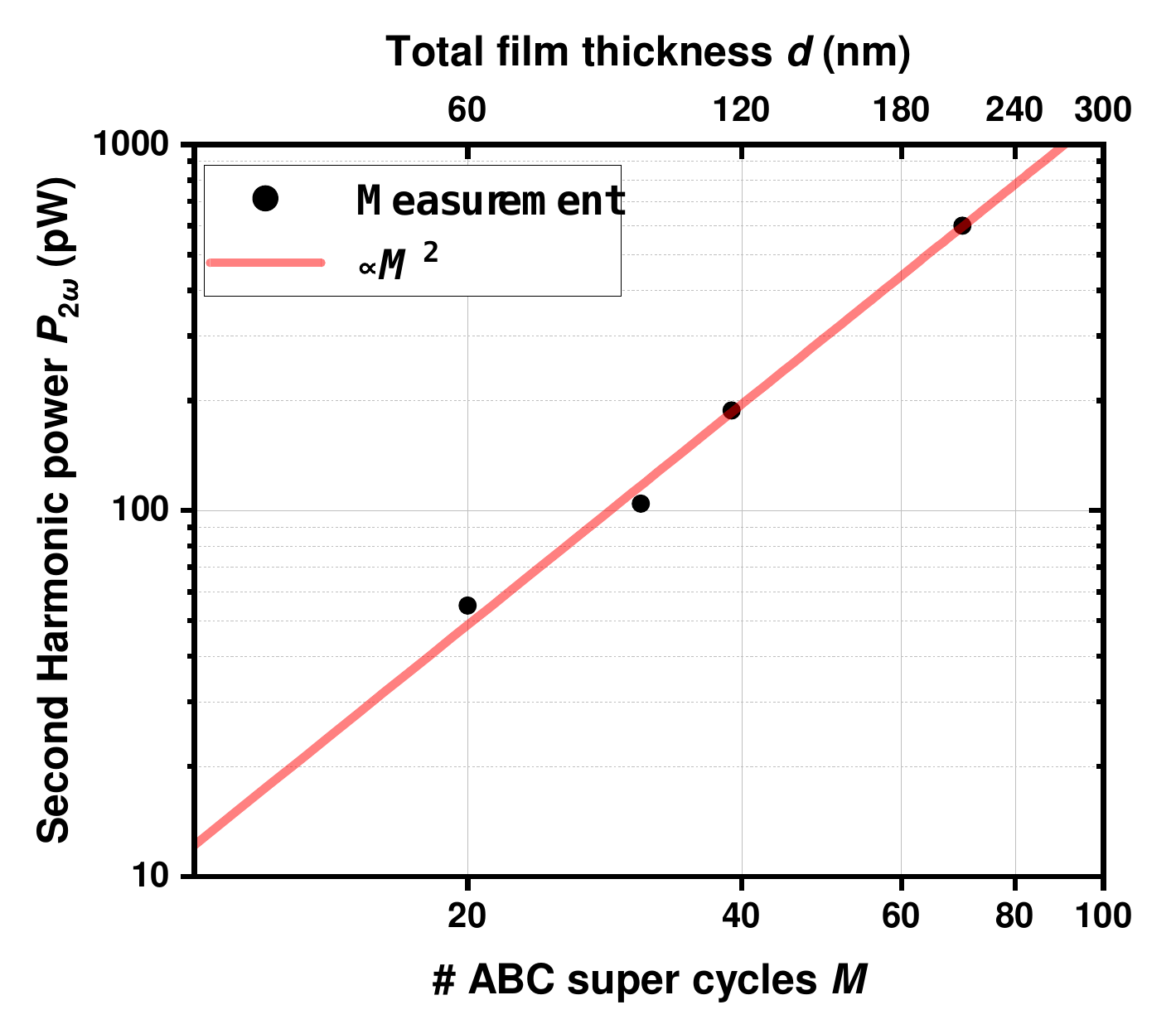}
\caption{Average second harmonic power scales with quadratic dependence on the film thickness and with the number of ABC super cycles $[\text{ABC}]\times M$. Both fundamental and SH are p-polarized with average fundamental power $P =0.5$ W. ABC period thickness is $t_{\textup{ABC}}=3.0$~nm for all samples. Measured at the angle of incidence $\vartheta$ where the global maxima occurs.}
\label{fig:Mpower}
\end{figure}

Several thicknesses were analyzed to clarify the properties at different deposited total thicknesses $d$. The second harmonic exhibits a quadratic relationship with the ABC super cycles $M$, similar to the classical bulk crystals, see Fig.\nobreakspace{}\ref{fig:Mpower}. Previously reported studies on ABC composites also identified the bulk-like SHG behavior~\cite{ABC-Alloatti, ABC-Clemmen}.

\begin{figure}[htbp]
\centering
\begin{subfigure}[t]{0.49\textwidth}
    \centering
    \includegraphics[scale=0.25]{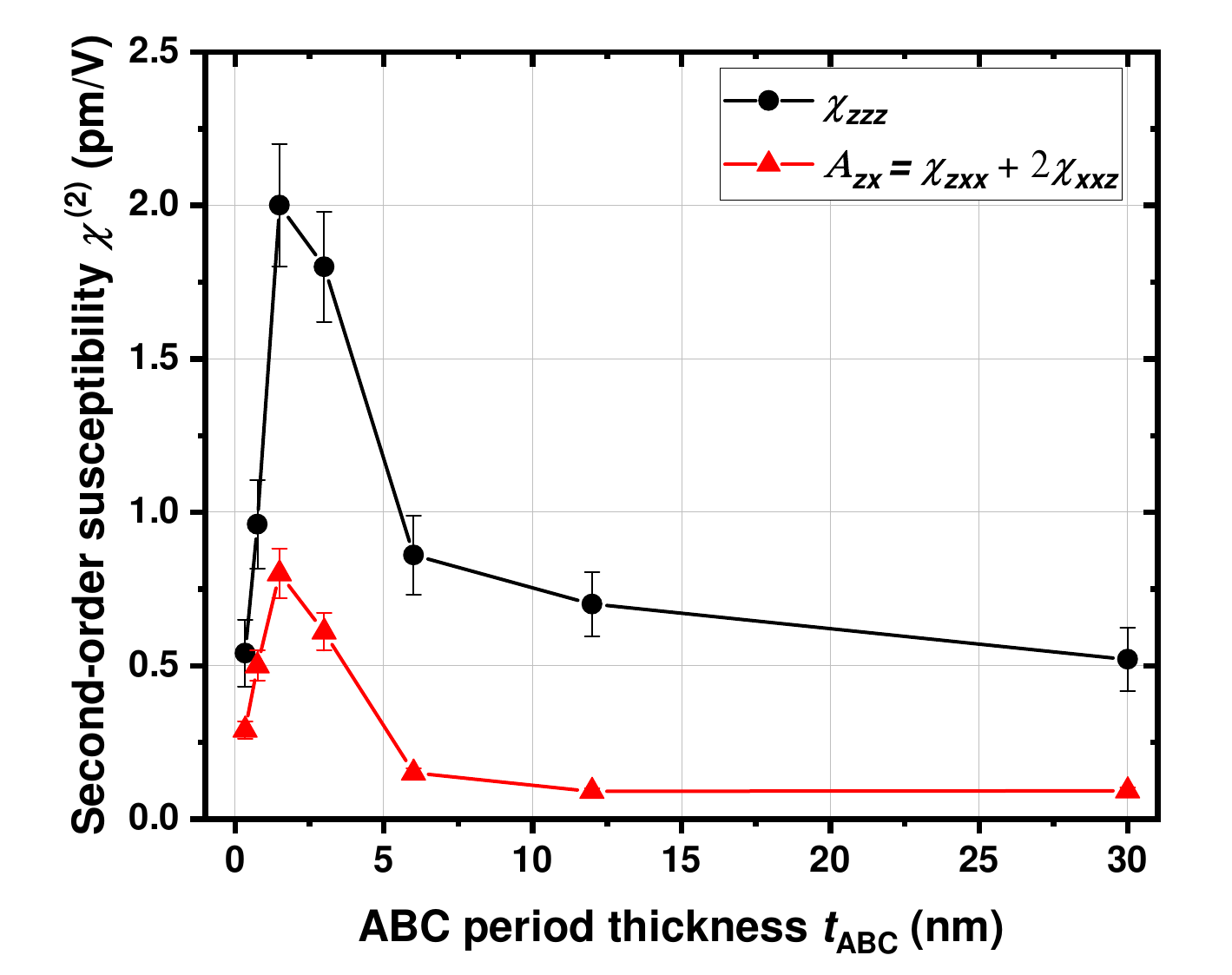}
    \caption{}
\label{fig:Chi2vsPeriod}
\end{subfigure}
\hfill
\begin{subfigure}[t]{0.49\textwidth}
    \centering
    \includegraphics[scale=0.25]{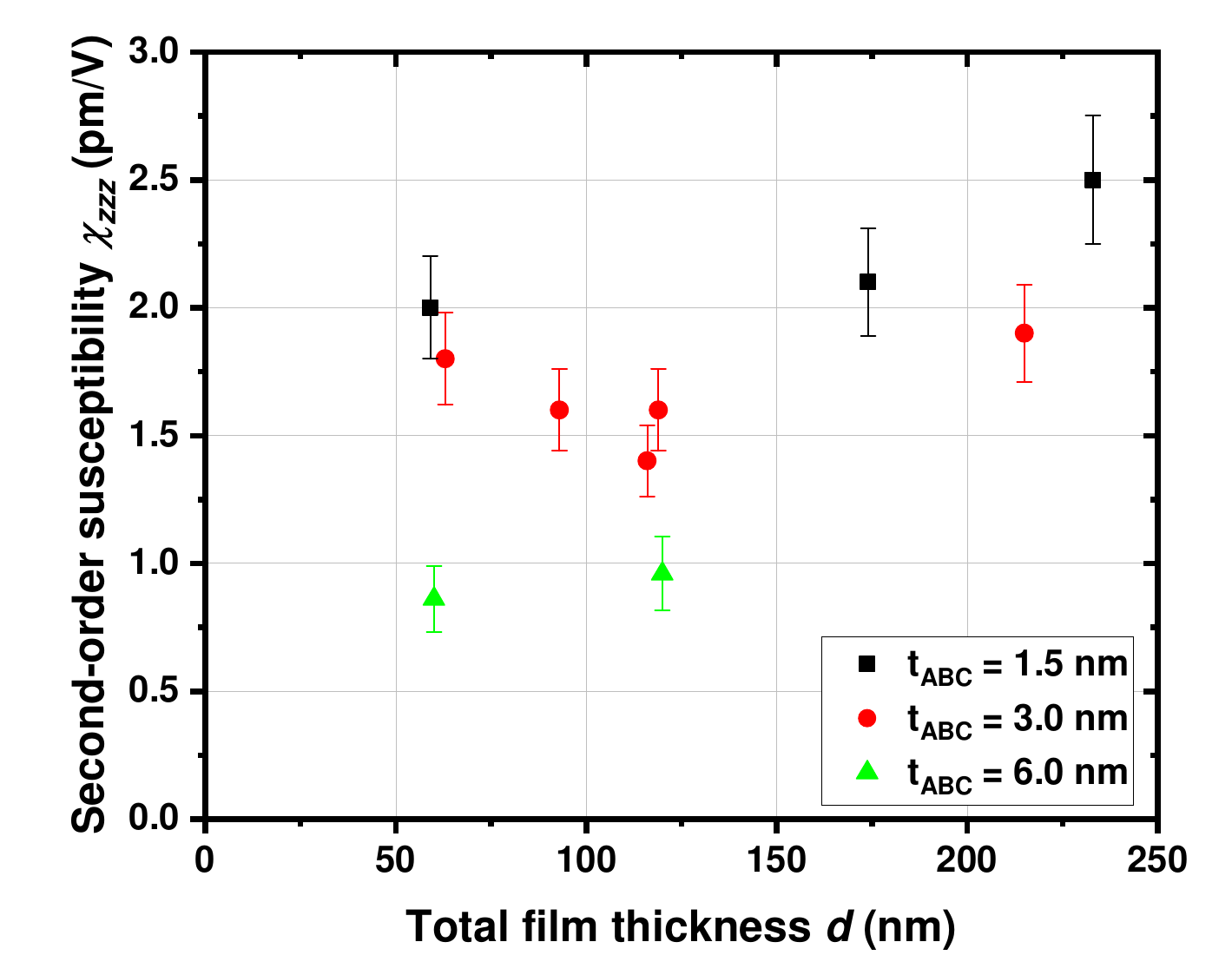}
    \caption{}
    \label{fig:Chi2vsThickness}
\end{subfigure}
\caption{(a) Analysis of second-order susceptibility $\chi_{zzz}^{}$ and $A_{zx}^{}$ as a function of ABC period thickness. (b) Investigating second-order susceptibility $\chi_{zzz}^{}$ across variations in ABC period and total film thickness.}
\label{fig:combined_STA_power}
\end{figure}

For each ABC period, second-order susceptibility is determined by fitting with the results shown in Fig. \ref{fig:Chi2vsPeriod}. The trend suggests an enhancement of both $\chi_{zzz}^{}$ and $A_{zx}^{}$ with decreasing ABC period thickness and achieving $\chi_{zzz}^{} = 2.0 \pm 0.2$\nobreakspace{}pm/V for $t_{\textup{ABC}}=1.5$\nobreakspace{}nm. For shorter periods, the values of $\chi_{zzz}^{}$ and $A_{zx}^{}$ decrease rapidly. The analysis also shows that the nonlinear susceptibility remains constant for total thicknesses from 60\nobreakspace{}nm to 233\nobreakspace{}nm for a period of 1.5\nobreakspace{}nm. A similar trend is also observed for periods of 3.0\nobreakspace{}nm and 6.0\nobreakspace{}nm in Fig. \ref{fig:Chi2vsThickness}.

Our best ABC compositions are compared in Table \ref{tab3} with other commonly used nonlinear crystals, as well as with previously reported results for ABC-type heterostructures in the literature. We achieve higher nonlinearity than previously reported values, and we attribute this improvement to higher quality thin film heterostructures, higher quality plasma source and ALD processes. For example, Alloatti et al. \cite{ABC-Alloatti} report difficulties with the HfO$_2$ process and overall structural inhomogeneity of the heterostructure. Moreover, in this systematic study we achieved much thinner, smoother, and still enclosed layers compared to the previous reports~\cite{ABC-Alloatti, ABC-Clemmen}. ABC period thicknesses $t_{\textup{ABC}}$ were 2.1\nobreakspace{}nm and 2.7\nobreakspace{}nm, compared to our best sample with 1.5\nobreakspace{}nm. Material selection may be partially responsible for the higher observed nonlinearity caused by the dielectric contrast between the layers.

\begin{table}[htbp]
\centering
\renewcommand{\heavyrulewidth}{0.1em} 
\renewcommand{\lightrulewidth}{0.05em} 
\setlength{\tabcolsep}{2pt} 
\caption{Comparing ABC-type heterostructures with other nonlinear monocrystals}
\scriptsize 
\renewcommand{\arraystretch}{0.8} 
\setlength{\aboverulesep}{0.2ex} 
\setlength{\belowrulesep}{0.2ex} 
\begin{tabular}{@{\hspace{2pt}}ccccccccc@{\hspace{2pt}}}
\toprule
\parbox[c]{1.2cm}{\centering Structure} 
& \parbox[c]{1.5cm}{\centering Material} 
& \parbox[c]{1.9cm}{\centering ABC material $n$ (-)} 
& \parbox[c]{1.5cm}{\centering $n_{\textup{o}}$/$n_{\textup{e}}$} 
& \parbox[c]{1.2cm}{\centering Point Group} 
& \parbox[c]{1.0cm}{\centering $\lambda$ (nm)} 
& \parbox[c]{1.2cm}{\centering $t_{\textup{ABC}}$ (nm)} 
& \parbox[c]{1.2cm}{\centering $d$ (pm/V)} 
& \parbox[c]{1.0cm}{\centering Ref.} \\
\toprule
Monocrystal & SiO$_2$ (Quartz) & - & 1.53/1.54 & $\textit{D}_{3}$ & 1064 & - & $d_\text{11}$ = 0.3 & \cite{lithiumbiobate,quartz-n} \\
\midrule
Monocrystal & BaB$_2$O$_4$ (BBO) & - & 1.67/1.54 & $\textit{C}_{3v}$ & 1064 & - & $d_\text{22}$ = 2.3 & \cite{lithiumbiobate,BBO-n} \\
\midrule
Monocrystal & LiNbO$_3$ (LN) & - & 2.23/2.16 & $\textit{C}_{3v}$ & 1064 & - & $d_\text{33}$ = -27 & \cite{lithiumbiobate,LN-n} \\
\midrule
ABC Amorphous & HfO$_2$/TiO$_2$/Al$_2$O$_3$ & 2.0/2.1/1.6 & - & $C_{\infty v}$ & 800 & 2.7 & $d_\text{33}$ = 0.4 & \cite{On-determination-SHG} \\
\midrule
ABC Amorphous & In$_2$O$_3$/TiO$_2$/Al$_2$O$_3$ & 2.2/2.1/1.6 & - & $C_{\infty v}$ & 800 & 2.1 & $d_\text{33}$ = 0.6 & \cite{On-determination-SHG} \\
\midrule
ABC Amorphous (S3) & SiO$_2$/TiO$_2$/Al$_2$O$_3$ & 1.45/2.33/1.62 & 1.77/1.75 & $C_{\infty v}$ & 1032 & 1.5 & $d_\text{33}$ = 1.0 & This work \\
\midrule
ABC Amorphous (S6) & SiO$_2$/TiO$_2$/Al$_2$O$_3$ & 1.45/2.33/1.62 & 1.79/1.71 & $C_{\infty v}$ & 1032 & 3.0 & $d_\text{33}$ = 0.9 & This work \\
\bottomrule
\end{tabular}
\smallskip
\footnotesize{By convention, $\chi_{}^{(2)}=2d$}
\label{tab3}
\end{table}

\begin{figure}[htbp]
\centering
\begin{subfigure}[c]{0.49\textwidth}
\centering
\includegraphics[scale=0.25]{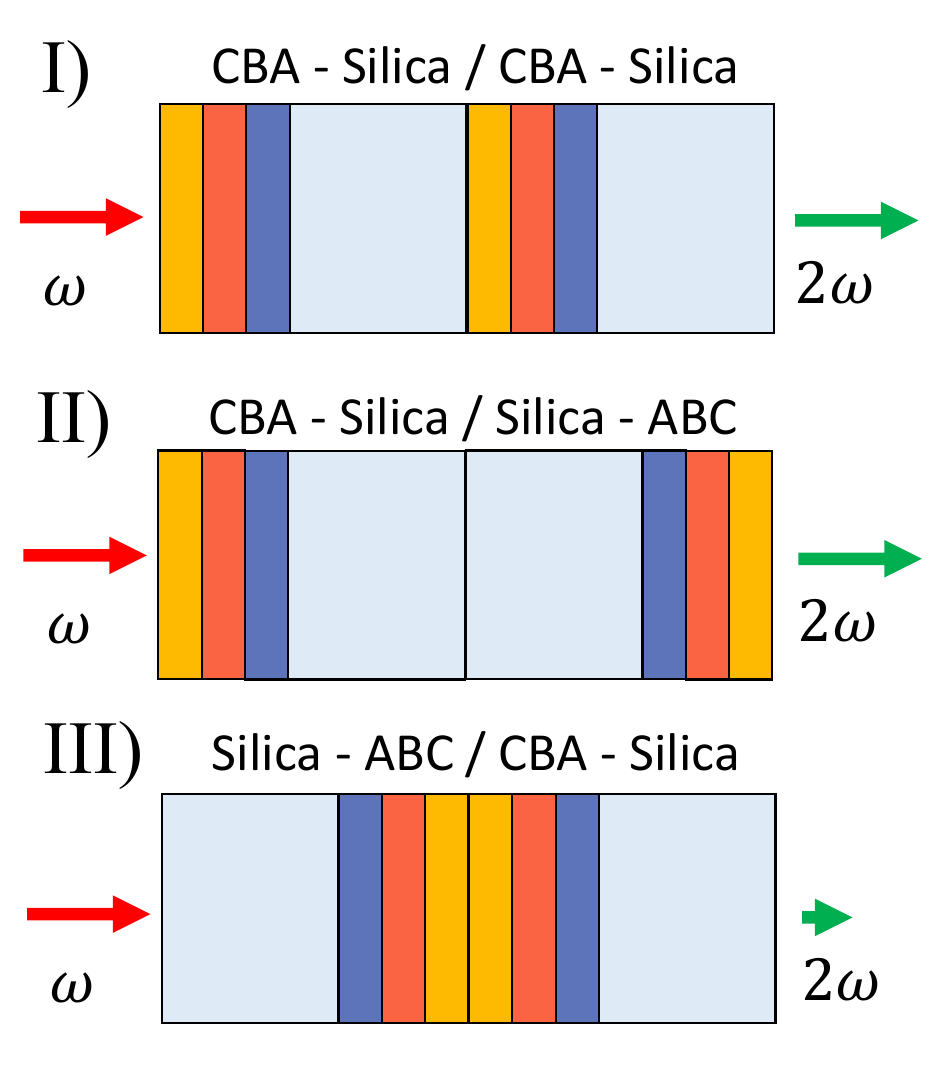}
\caption{}
\label{2}
\end{subfigure}
\hfill
\begin{subfigure}[c]{0.49\textwidth}
\centering
\includegraphics[scale=0.25]
{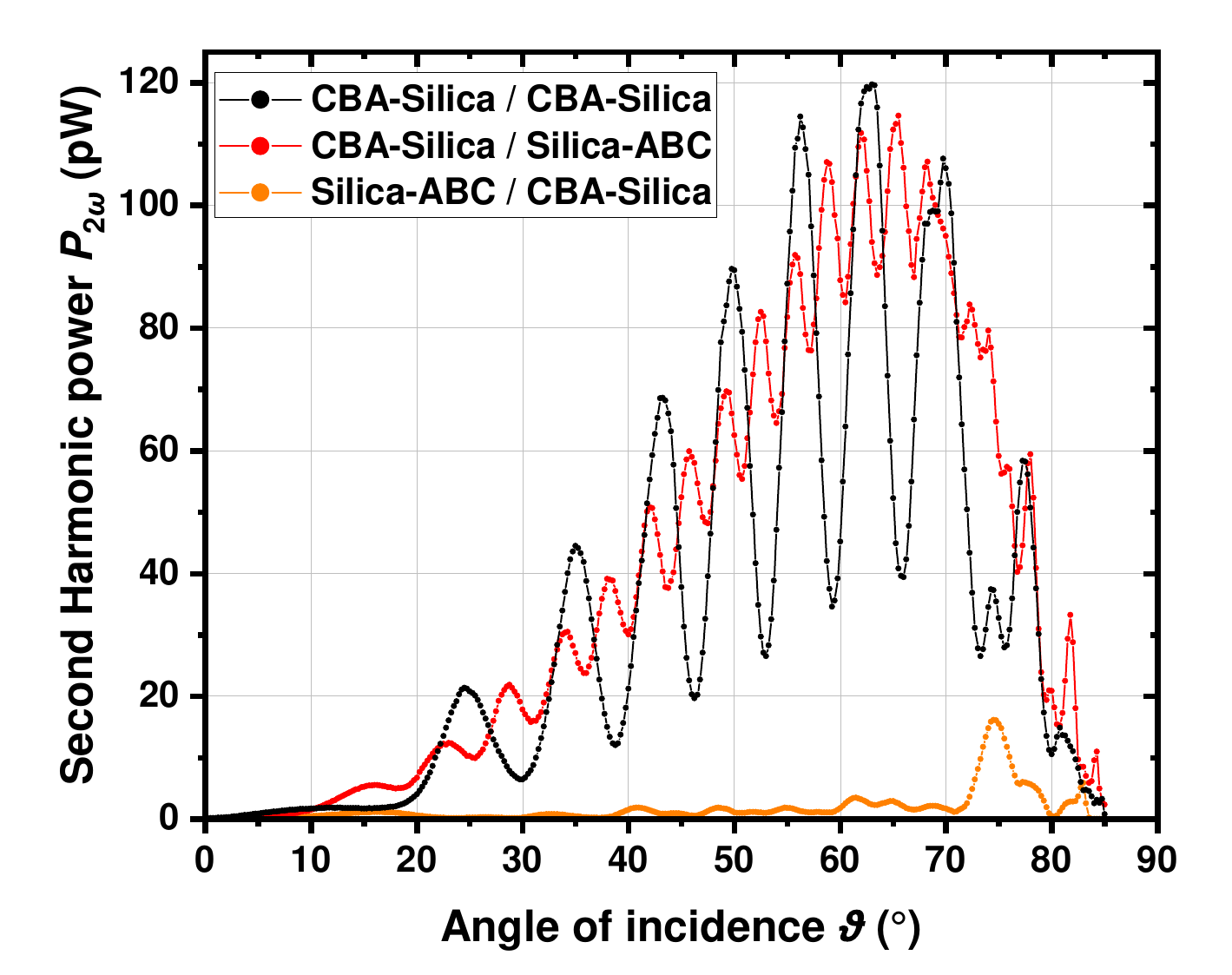}
\caption{}
\end{subfigure}
\caption{(a) Showing each stacking sequence for the two identical samples with total deposited ABC stack thickness of 60\nobreakspace{}nm and thickness of the fused silica substrate is 1\nobreakspace{}mm. (b) Showing angle-dependent measurements for three distinct stacking sequences of the ABC sample, both for p-polarized fundamental and second harmonic. Depending on the stacking sequence configuration, the SH response varies significantly. Both FF with an average pump power $P=0.5$\nobreakspace{}W and SH are p-polarized.}
\label{fig:CBA}
\end{figure}

To further clarify the origin of the effective bulk nonlinearity~\cite{ABC-Clemmen}, two identical ABC samples S3 were placed on top of each other. First, two samples with stacking sequence I) CBA...CBA-Silica/CBA...CBA-Silica and sequence II) CBA...CBA-Silica/Silica-ABC...ABC were investigated, see Fig. \ref{fig:CBA}. Both stacking sequences produce almost the same maximum SHG power. However, the number of interference fringes is increased for sequence II), and the contrast of the fringes is reduced. In this case, the temporal walk-off effect is more significant because the optical thickness of the substrates between the heterostructures is twice as large. However, when the stacking sequence III) Silica-ABC...ABC/CBA...CBA-Silica is applied, the SHG signal almost vanishes. This result suggests that the overall centrosymmetry of the stack is restored. However, SH waves must be generated in each ABC and CBA layer; therefore, they are generated in each stack with an opposite phase and destructively interfere. This leads to the conclusion that the two susceptibilities of the ABC and CBA layers must have an opposite sign, for example, $\chi_{\textup{s}}^{\textup{ABC}}=-\chi_{\textup{s}}^{\textup{CBA}}$. The polarity flipping of the second-order susceptibility by reversing the growing order was also demonstrated in semiconductor heterostructures based on quantum-engineered intersubband transitions, where the polarity is flipped by spatial inversion in the growth sequence \cite{chi2flipping}, similar to our case. However, the ABC and CBA sequences can differ under real conditions. For example, material A can grow on material B at a different rate and quality than material B on A. It is yet to be determined whether all growth permutations ABC, ACB, BAC, BCA, CAB, and CBA result in equivalent stack characteristics.

\section{Conclusion}
We demonstrated an enhancement of the effective bulk second-order susceptibility in ABC-type heterostructures by optimizing the thickness of the ABC period, achieving a dominant tensor component of $\chi_{zzz} = 2.0 \pm 0.2$\nobreakspace{}pm/V at $t_{\text{ABC}} = 1.5$\nobreakspace{}nm, based on the second-order nonlinear response at the interfaces. The observed non-linearity enhancement stems from the higher density of interfaces, symmetry breaking in the ABC stacks, and dielectric contrast between the materials. The nonlinearity could be further enhanced if the layers remain distinct and non-intermixed at the range of 0.5\nobreakspace{}nm, for example, by minimizing surface and interface roughness through deposition on ultra-smooth substrates. Extremely thin and homogeneous layers are needed. The evolution of the roughness inside the ALD multilayer stack is the result of both the replicated substrate roughness and the intrinsic film roughness. The coating material and process conditions can influence the latter. This may have the potential for further improvements, particularly in increasing layer density~\cite{schroder2010angle}. Among other techniques, ALD has a high potential to provide such films for future advanced nonlinear photonic application. Experimental results indicate that the optimal strategy to maximize second-order susceptibility is to incorporate the greatest number of ABC layers per unit thickness, while ensuring physical separation to maintain structural integrity. However, the layer thickness is ultimately constrained by the limit of a single atomic layer. Therefore, future efforts should also prioritize the identification of a more suitable material composition, which remains challenging and elusive. Additionally, ABC nanolaminates have potential for near-ultraviolet applications, leveraging their low absorption edge when constituent materials are carefully selected. Polarity flipping, combined with form birefringence, enables phase matching in bulk form-birefringent metamaterials~\cite{formBirometamaterial}. We envision that ABC-type heterostructures will open new avenues for nonlinear optics through integration into nanophotonic waveguides to efficiently access the dominant tensor component $\chi_{zzz}$~\cite{ABC-waveguide}, and current efforts focus on integrating the heterostructures into nonlinear metasurfaces and other photonic platforms~\cite{ABC-overcoating,ABC-metasrufaceplasmonic}.

\begin{backmatter}
\bmsection{Funding}
This research received funding from the Deutsche Forschungsgemeinschaft (DFG, German Research Foundation) through the International Research Training Group (IRTG) 2675 “Meta-ACTIVE” (project number 437527638) and the Collaborative Research Center (CRC) 1375, (project number 398816777) "NOA". Additional support was provided by the Fraunhofer Internal Programs under Grant No. SME 431 40-04871. We also gratefully acknowledge funding from the Federal Ministry of Research, Technology, and Space under Grant No. 13N16897 (GRADIENT) and 13F1000A (PriFusio). The authors also thank the European Regional Development Fund and the State of Brandenburg for the Themis Z microscope (part of the Potsdam Imaging and Spectral Analysis (PISA) facility).

\bmsection{Disclosures}
The authors declare no conflict of interest.

\bmsection{Data Availability}
Data underlying the results presented in this paper are not publicly available at this time but may be obtained from the authors upon reasonable request.

\bmsection{Supplemental document}
See Supplement 1 for additional details of the nonlinear characterization.


\bibliography{sample}


\end{backmatter}
\end{document}